\documentclass{article}

\scrollmode
\usepackage{amsmath}
\usepackage{amsfonts}
\usepackage{amssymb}
\usepackage{latexsym}
\usepackage{stmaryrd}
\usepackage{array}
\usepackage{exscale}
\newcommand{\nc}{\newcommand}
\newcommand{\ol}{\overline}

\newcommand{\es}{\emptyset}
\newcommand{\sm}{\setminus}
\newcommand{\ve}{\varepsilon}
\newcommand{\vp}{\varphi}
\newcommand{\bw}{\bigwedge}
\newcommand{\bv}{\bigvee}
\newcommand{\bc}{\bigcup}

\newcommand{\Lra}{\Leftrightarrow}

\newcommand{\Ra}{\Rightarrow}

\newcommand{\ra}{\rightarrow}

\newcommand{\lra}{\leftrightarrow}

\newcommand{\sse}{\subseteq}

\newcommand{\spe}{\supseteq}
\newcommand{\fa}{\forall}
\newcommand{\ex}{\exists}
\newcommand{\mr}{\mathrm}
\newcommand{\mc}{\mathcal}
\newcommand{\mf}{\mathfrak}

\newcommand{\DMO}{\DeclareMathOperator}
\newcommand{\DST}{\displaystyle}

\newcommand{\ZZ}{\mathbb{Z}}
\newcommand{\NN}{\mathbb{N}}
\newcommand{\NNZ}{\NN_0}

\newcommand{\RR}{\mathbb{R}}

\newcommand{\aru}{\ar @{-}} 
\usepackage{listings}
\newcommand{\inl}[1]{\lstinline$#1$}
\newcommand{\und}{{\:\wedge\:}} 
\newcommand{\oder}{{\:\vee\:}} 
\newcommand{\mb}{{\:|\:}} 
\newcommand{\set}[1]{\{ #1 \}}
\newcommand{\setb}[1]{\big \{ \, #1 \, \big \}}
\nc{\simlvi}[1]{\!\sim_{#1}}
\nc{\apprel}[3]{{#1}(#2)_{(#3)}} 
\nc{\cmpli}[1]{\complement^1_{#1}} 
\nc{\cmplzi}[1]{\complement^0_{#1}} 
\nc{\cmplzoi}[1]{\complement^*_{#1}} 

\nc{\zf}{\mr{ZF}}
\nc{\zfmf}{\zf^0} 
\nc{\zfc}{\mr{ZFC}}
\nc{\zfcmf}{\zfc^0} 
\nc{\bst}{\mr{BST}} 
\newcommand{\tb}[2]{\set{#1, \dots, #2}} 
\providecommand{\abs}[1]{\lvert #1 \rvert} 
\providecommand{\norm}[1]{\lVert #1 \rVert} 
\providecommand{\inprod}[1]{\left\langle #1 \right\rangle} 
\newcommand{\trans}[1]{#1^{\hspace{0.05em}\mr{t}}} 
\makeatletter
\DeclareRobustCommand{\genericinterval}[2]{%
  \@ifstar{\genericinterval@star{#1}{#2}}{\genericinterval@nostar{#1}{#2}}}
\newcommand{\genericinterval@star}[4]{\mathopen{}\mathclose{\left#1#3,#4\right#2}}
\newcommand{\genericinterval@nostar}[4]{\mathopen{#1}#3,#4\mathclose{#2}}

\makeatother
\nc{\untit}[2]{{#1}^{#2 \downarrow}} 
\nc{\obit}[2]{{#1}^{#2 \uparrow}} 
\nc{\inzEKi}[1]{\mc{I}^{\mr{V}}_{#1}}

\nc{\inzKEi}[1]{\mc{I}^{\mr{E}}_{#1}}

\nc{\adjEi}[1]{\mc{A}^{\mr{V}}_{#1}}

\nc{\BD}[1]{{#1}\text{-}\mr{BD}}

\nc{\konv}[2]{{#1}[{#2}]} 
\nc{\actpres}[1]{\Phi_{#1}} 
\nc{\Prim}{\mc{PR}} 

\nc{\sselr}{\sse^{\mapsto}}
\nc{\sserl}{\sse^{\mapsfrom}}
\nc{\spelr}{\spe^{\mapsto}}
\nc{\sperl}{\spe^{\mapsfrom}}
\nc{\ball}[1]{\mr{B}^{#1}} 
\nc{\oball}[1]{\breve{\mr{B}}^{#1}} 
\nc{\pball}[1]{\dot{\mr{B}}^{#1}} 
\nc{\prr}[1]{\dot{\RR}^{#1}} 
\nc{\sph}[1]{\mr{S}^{#1}} 
\nc{\ssim}[1]{s\sigma_{#1}} 
\nc{\koerper}[1]{\norm{#1}}
\nc{\Ccovdim}{\mc{CD}}
\nc{\Cinddim}{\mc{SID}}

\nc{\CInddim}{\mc{LID}}

\DeclareMathOperator{\diffop}{D} 
\DeclareMathOperator*{\diffoplimit}{D} 
\nc{\diffopc}[1]{\sideset{_{#1}}{}\diffoplimit} 
\nc{\diffopp}[1]{\diffop_{#1}} 
\nc{\diffopcp}[2]{\sideset{_{#2}}{_{#1}}\diffoplimit} 
\nc{\meanH}[2]{\mf{M}_{#1,#2}} 
\nc{\emean}[2]{\mf{M}_{\exp_{#1},#2}} 
\DeclareMathOperator{\mor}{Mor}
\DeclareMathOperator{\Hom}{Hom} 
\nc{\autoerw}[1]{\mr{Aut}^{#1}} 
\nc{\komma}[2]{(#1 \downarrow #2)} 
\nc{\Kmat}{\mf{MAT}} 
\nc{\Khmat}{\mf{HMAT}} 
\nc{\homfun}[1]{\mor_{#1}(-_1,-_2)} 
\nc{\homfunae}[1]{\mor_{#1}(-_1)} 
\nc{\homfunbe}[1]{\mor_{#1}(-_2)} 
\nc{\homfunxy}[3]{\mor_{#1}(#2(-_1), #3(-_2))}
\nc{\homfunx}[2]{\mor_{#1}(#2(-_1), -_2)}
\nc{\homfuny}[2]{\mor_{#1}(-_1, #2(-_2))}
\nc{\homfuna}[2]{\mor_{#1}(#2, -)} 
\nc{\homfunb}[2]{\mor_{#1}(-, #2)} 
\nc{\hhomfuna}[2]{\Hom_{#1}(#2, -)} 
\nc{\hhomfunb}[2]{\Hom_{#1}(-, #2)} 
\newcommand{\Va}{\mc{V\hspace{-0.1em}A}}

\newcommand{\Lit}{\mc{LIT}}
\newcommand{\Cl}{\mc{CL}}
\newcommand{\Cls}{\mc{CLS}}

\newcommand{\Pcls}[1]{#1\mbox{--}\Cls}

\newcommand{\Pass}{\mc{P\hspace{-0.32em}ASS}}
\newcommand{\Tass}{\mc{T\hspace{-0.35em}ASS}}
\newcommand{\Sat}{\mc{SAT}}

\newcommand{\Usat}{\mc{USAT}}

\newcommand{\Musat}{\mc{M\hspace{0.8pt}U}} 
\newcommand{\Musati}[1]{\Musat_{\!#1}} 
\newcommand{\Smusat}{\mc{S}\Musat} 
\newcommand{\Smusati}[1]{\Smusat_{\!#1}}

\nc{\Clsoo}{\Cls^{1,1}} 
\DeclareMathOperator{\lit}{lit}
\DeclareMathOperator{\var}{var}

\newcommand{\Clash}{\mc{HIT}} 

\newcommand{\Uclash}{\mc{U}\Clash} 
\newcommand{\Uclashi}[1]{\Uclash_{\!\!#1}}

\newcommand{\Ho}{\mc{HO}} 

\DeclareMathOperator{\res}{\diamond} 

\DMO{\premr}{F} 
\DMO{\concr}{C} 
\DMO{\allcr}{\widehat{F}} 
\DMO{\semspace}{ss} 
\DMO{\treespace}{ts} 

\DeclareMathOperator{\hardness}{hd}
\DMO{\phardness}{phd} 
\DMO{\whardness}{whd} 
\DMO{\hts}{hs} 
\newcommand{\php}{\mathrm{PHP}}
\newcommand{\ephp}{\mathrm{EPHP}} 
\newcommand{\pab}[1]{\langle #1 \rangle}
\newcommand{\pao}[2]{\langle #1 \ra #2 \rangle}

\nc{\bth}[1]{\langle{#1}\rangle} 
\DMO{\rsub}{r_S} 
\DMO{\rk}{r} 
\DMO{\rki}{r_{\infty}} 
\nc{\rslur}{\xrightarrow{\text{SLUR}}} 
\nc{\rslurs}{\rslur_{\!*}} 
\DMO{\slur}{slur} 
\nc{\Slur}{\mc{SLUR}} 
\nc{\rkslur}[1]{\xrightarrow{\text{SLUR}_{#1}}} 
\nc{\rkslurs}[1]{\rkslur{#1}_{\!*}} 
\nc{\Altsluri}[1]{\Slur(#1)}
\nc{\Altslurstari}[1]{\Slur\text{\textasteriskcentered}(#1)}
\nc{\Canoni}[1]{\mr{CANON}(#1)}
\nc{\rkslurstar}[1]{\xrightarrow{\text{SLUR\textasteriskcentered}#1}} 
\nc{\rkslursstar}[1]{\rkslurstar{#1}_{\!*}} 
\DMO{\slurstar}{\slur\!\text{\textasteriskcentered}}
\nc{\Urefc}{\mc{UC}}
\nc{\Propc}{\mc{PC}}
\nc{\Wrefc}{\mc{WC}} 
\DeclareMathOperator{\wid}{wid} 

\DeclareMathOperator{\vdeg}{vd} 
\DeclareMathOperator{\minvdeg}{\mu\!\vdeg} 
\DMO{\varmvd}{\var_{\minvdeg}} 
\DMO{\nfc}{fc} 
\DMO{\maxnfc}{\nu\!\nfc} 
\nc{\svbf}{\mc{VB}} 
\nc{\svbfs}{\mc{VB}^*} 
\DMO{\potp}{pp} 
\DMO{\potprec}{NM} 
\DMO{\minnonmer}{\mu{}nM} 
\DMO{\varsing}{\var_s} 
\DMO{\varosing}{\var_{1s}} 
\DMO{\varnosing}{\var_{\neg1s}} 
\nc{\Musatns}{\Musat'} 
\nc{\Musatnsi}[1]{\Musati{#1}'}
\nc{\Smusatns}{\Smusat'} 
\nc{\Smusatnsi}[1]{\Smusati{#1}'}
\nc{\Uclashns}{\Uclash'} 
\nc{\Uclashnsi}[1]{\Uclashi{#1}'}
\nc{\tsdp}{\xrightarrow{\text{sDP}}}
\nc{\tsdps}{\tsdp_{\!*}}
\nc{\tosdp}{\xrightarrow{\text{1sDP}}}
\nc{\tosdps}{\tosdp_{\!*}}
\DMO{\sdp}{sDP} 
\DMO{\osdp}{sDP_1} 
\nc{\cflmusat}{\mc{CF}\Musat} 
\nc{\cflmusati}[1]{\mc{CF}\Musati{#1}}
\nc{\cflimusat}{\mc{CFI}\Musat} 
\DMO{\sNF}{sNF} 
\DMO{\eqp}{eqp} 
\DMO{\sgp}{sp} 
\DMO{\singind}{si} 
\DMO{\osingind}{si_1} 
\DMO{\shyp}{svh} 
\DMO{\sdph}{ssh} 
\DMO{\msdph}{mss} 
\DMO{\osdph}{ssh_1} 
\DMO{\mosdph}{mss_1} 
\DMO{\mps}{mps} 
\DMO{\purec}{puc} 
\DMO{\doping}{D}
\DeclareMathOperator{\primec}{prc} 
\nc{\glue}[4]{\mr{glue}((#1,#2), (#3,#4))} 
\DMO{\fvdglue}{\boxplus} 
\nc{\gluea}[3]{#1 \boxplus_{#3} #2} 
\DMO{\frl}{fl} 
\nc{\Con}{\mr{Con}}
\nc{\Log}{\mr{Log}}
\nc{\Lin}{\mr{Lin}}
\nc{\Pol}{\mr{Pol}}
\nc{\ExL}{\mr{ExL}}
\nc{\ExP}{\mr{ExP}}
\nc{\CTime}{\mr{CTime}}
\nc{\CSpace}{\mr{CSpace}}
\nc{\LTime}{\mr{LTime}}
\nc{\LSpace}{\mr{L}}
\nc{\NLSpace}{\mr{NL}}
\nc{\LinTime}{\mr{LinTime}}
\nc{\LinSpace}{\mr{LinSpace}}
\nc{\PTime}{\mr{P}}
\nc{\PSpace}{\mr{PSpace}}
\nc{\Np}{\mr{NP}}
\nc{\Conp}{\text{coNP}}
\nc{\NPSpace}{\mr{NPSpace}}
\nc{\CoNPSpace}{\mr{coNPSpace}}
\nc{\ELTime}{\mr{ELTime}}
\nc{\ELSpace}{\mr{ELSpace}}
\nc{\EPTime}{\mr{EPTime}}
\nc{\EPSpace}{\mr{EPSpace}}
\nc{\NEPTime}{\mr{NEPTime}}
\nc{\polydelta}[1]{\Delta_{#1}^{\mr P}}
\nc{\polypi}[1]{\Pi_{#1}^{\mr P}}
\nc{\polysigma}[1]{\Sigma_{#1}^{\mr P}}
\nc{\Ph}{\mr{PH}}

\nc{\Dp}{D^P}
\nc{\PllC}[2]{{\text{$\mr{PT}$/$\mr{WK}$}(#1, #2)}} 
\nc{\Nc}{\mr{NC}}
\nc{\Nci}[1]{\Nc^{#1}}
\nc{\Ac}{\mr{AC}}
\nc{\Aci}[1]{\Ac^{#1}}
\nc{\pmodpoly}{P / \mathrm{poly}}
\nc{\Wh}[1]{\mr{W}[#1]} 
\nc{\Rl}{\mr{RL}}
\nc{\coRl}{\mr{coRL}}
\nc{\Rp}{\mr{RP}}
\nc{\coRp}{\mr{coRP}}
\nc{\Zpp}{\mr{ZPP}}
\nc{\Bpp}{\mr{BPP}}
\nc{\Pp}{\mr{PP}}
\nc{\Reach}{\mr{STCON}} 
\nc{\Undreach}{\mr{USTCON}} 
\nc{\Pcol}[2]{\mr{COL}(#1,#2)} 
\nc{\Pscol}[2]{\mr{SCOL}(#1,#2)} 
\nc{\Psorcol}[2]{\mr{SORCOL}(#1,#2)} 
\nc{\Mss}{\mr{MSS}}
\nc{\Key}{\mr{KEY}}
\nc{\Keyi}[1]{\Key_{\!#1}}
\nc{\Nbmss}{N_{\mr{bm}}} 
\nc{\Nbkey}{N_{\mr{bk}}} 
\nc{\Rnb}{N_{\mr{b}}}
\nc{\Rnk}{N_{\mr{k}}}
\nc{\Rnr}{N_{\mr{r}}}

\nc{\Byte}{\mr{B}[8]}
\nc{\QByte}{\mr{B}[4,8]}
\nc{\KByte}{\mc{B}} 
\nc{\RQByte}{\mc{QB}} 

\nc{\ramz}[3]{\mr{ram}_{#1}^{#2}(#3)} 
\nc{\waez}[2]{\mr{vdw}_{#1}(#2)} 
\nc{\gtz}[2]{\mr{grt}_{#1}(#2)} 
\nc{\pdwaez}[2]{\mr{vdw}_{#1}^{\mr{pd}}(#2)} 
\nc{\absfeh}[1]{\delta_{#1}} 
\nc{\relfeh}[1]{\ve_{#1}} 
\usepackage{theorem} 
\usepackage[driverfallback=hypertex]{hyperref}
\newtheorem{defi}{Definition}[section]
\newtheorem{lem}[defi]{Lemma}
\newtheorem{thm}[defi]{Theorem}
\newtheorem{corol}[defi]{Corollary}

\newtheorem{conj}[defi]{Conjecture}

\newtheorem{examp}[defi]{Example}

\theorembodyfont{\rmfamily}

\theorembodyfont{}
\newenvironment{prf}{\noindent\textbf{Proof:}\;}{\par\noindent\ignorespacesafterend}

\newcommand{\Qed}{\hfill $\square$}
\newcounter{dDef} 

\newcounter{dLem} 

\newcounter{dThm} 

\newcounter{dPro} 

\newcounter{Beispielzaehler}

\nc{\bm}{\boldmath}
\nc{\bmm}[1]{\mbox{\bm$\DST #1$}}
\nc{\mi}[1]{\bmm{\mathrm{(#1):}} \quad}

\usepackage[active]{srcltx}
\usepackage{a4}
\usepackage[all,poly]{xy}
\usepackage{float}
\usepackage{bussproofs}

\DMO{\thardness}{thd}

\DMO{\EUrefc}{\exists\,\Urefc}

\DMO{\twidth}{tw}

\begin{document}

\title{On SAT representations of XOR constraints}

\author{Matthew Gwynne and Oliver Kullmann\\
  \qquad {\small \url{http://cs.swan.ac.uk/~csmg/}} \qquad
  {\small \url{http://cs.swan.ac.uk/~csoliver}}\\
  \href{http://www.swan.ac.uk/compsci/}{Computer Science Department}\\
  \href{http://www.swan.ac.uk/}{Swansea University}\\
  Swansea, UK
}

\maketitle

\begin{abstract}
  We study the problem of finding good CNF-representations $F$ of systems of linear equations $S$ over the two-element field, also known as systems of XOR-constraints $x_1 \oplus \dots \oplus x_k = \ve$, $\ve \in \set{0,1}$, or systems of parity-constraints. The number of equations in $S$ is $m$, the number of variables is $n$. These representations are used as parts of SAT problems $F^* \supset F$, such that $F$ has ``good'' properties for SAT solving in the context of $F^*$; here $F^*$ may for example represent the problem of finding the key for a cryptographic cipher. The basic quality criterion is ``arc consistency'' (AC), that is, for every partial assignment $\vp$ to the variables of $S$, all assignments $x_i = \ve$ forced by $\vp$ are determined by unit-clause propagation on the result $\vp * F$ of the application.

  We show there is no AC-representation of polynomial size for arbitrary $S$. We use the lower bound on monotone circuits for monotone span programs from \cite{BabaiGalWigderson1999MonoteSpanPrograms}, and we show a close relation between monotone circuits and AC-representations, based on the work in \cite{BKNW2009CircuitComplexity}. We then turn to constructing good representations. We analyse the basic translation $F = X_1(S)$, which translates each constraint on its own, by splitting up $x_1 \oplus \dots \oplus x_k$ into sums $x_1 \oplus x_2 = y_2, y_2 \oplus x_3 = y_3, \dots, y_{k-1} \oplus x_k = \ve$, introducing auxiliary variables $y_i$. We show that $X_1(S^*)$, where $S^*$ is obtained from $S$ by considering all derived equations, is an AC-representation of $S$. The derived equations are obtained by adding up the equations of all sub-systems $S' \sse S$. There are $2^m$ such $S'$, and computing an AC-representation is fixed-parameter tractable (fpt) in the parameter $m$, improving \cite{LaitinenJunttilaNiemelae2013Parity}, which showed fpt in $n$.

  To obtain stronger representations, instead of mere AC we consider the class $\Propc$ of propagation-complete clause-sets, as introduced in \cite{BordeauxMarquesSilva2012KnowledgeCompilation}. The stronger criterion is $F \in \Propc$, which requires for \emph{all} partial assignments, possibly involving also the auxiliary (new) variables in $F$, that forced assignments can be determined by unit-clause propagation. Using ``propagation hardness'' $\phardness(F) \in \NNZ$ as introduced in \cite{GwynneKullmann2012SlurJ}, we have $F \in \Propc \Lra \phardness(F) \le 1$. We show that $X_1$ applied to a single equation ($m=1$) yields a translation in $\Propc$, i.e., $\phardness(X_1(S)) \le 1$. Then we study $m=2$. Now $S^*$ has two equations more, and $X_1(S^*)$ is an AC-representation, but the ``distance'' to $\Propc$ is arbitrarily high, i.e., $\phardness(X_1(S^*))$ is unbounded (using results from \cite{BeyersdorffGwynneKullmann2013PHPER}). We show two possibilities to remedy this (for $m=2$). On the one hand, if instead of unit-clause propagation we allow (arbitrary) resolution with clauses of length at most $3$ (i.e., $3$-resolution), and only require refutation of inconsistencies after (arbitrary, partial) instantiations, then even just $X_1(S)$ suffices. On the other hand, with a more intelligent translation, which avoids duplication of equivalent auxiliary variables $y_i$, we obtain a (short) representation in $\Propc$. We conjecture that also the general case can be handled this way, that is, computing a representation $F \in \Propc$ of $S$ is fpt in $m$.
\end{abstract}

\tableofcontents

\section{Introduction}
\label{sec:intro}

Recall that the two-element field $\ZZ_2$ has elements $0,1$, where addition is XOR, which we write as $\oplus$, while multiplication is AND, written $\cdot$. A linear system $S$ of equations over $\ZZ_2$, in matrix form $A \cdot x = b$, where $A$ is an $m \times n$ matrix over $\set{0,1}$, with $m$ the number of equations, $n$ the number of variables, while $b \in \set{0,1}^m$, yields a boolean function $f_S$, which assigns $1$ to a total assignments of the $n$ variables of $S$ iff that assignment is a solution of $S$. The task of ``good'' representations of $f_S$ by conjunctive normal forms $F$ (clause-sets, to be precise), for the purpose of SAT solving, shows up in many applications, for example cryptanalysing the Data Encryption Standard and the MD5 hashing algorithm in \cite{BardCourtois2007AlgebraicDES}, translating Pseudo-Boolean constraints to SAT in \cite{Een2006Translating}, and in roughly $1$ in $6$ benchmarks from SAT 2005 to 2011 according to \cite{LaitinenJunttilaNiemelae2012Parity} (see Subsection \ref{sec:litrev} for an overview on the literature).

The basic criterion for a good $F$ is ``arc-consistency'', which we write as ``AC-representation'' (similar to ``$\Propc$-representation'', explained later); in Subsection \ref{sec:introAC} we review and discuss this terminology. For an arbitrary boolean function $f$, a CNF-representation $F$ of $f$ is AC if for every partial assignment $\vp$ to the variables of $f$, when applying the partial assignment to $F$, i.e., performing $F \leadsto \vp * F$, and then performing unit-clause propagation, which we write as $\rk_1$, the result $F' := \rk_1(\vp * F)$ has no forced assignments anymore, that is, for every remaining variable $v$ and $\ve \in \set{0,1}$ the result $\pao v{\ve} * F'$ of assigning $\ve$ to $v$ in $F'$ is satisfiable.

\subsection{The lower bound}
\label{sec:introlowb}

We show that there is no polynomial-size AC-representation of arbitrary linear systems $S$ (Theorem \ref{thm:xorclsrel}). To show this, we apply the lower bound on monotone circuit sizes for monotone span programs (msp's) from \cite{BabaiGalWigderson1999MonoteSpanPrograms}, by translating msp's into linear systems. An msp computes a boolean function $f(x_1,\dots,x_n) \in \set{0,1}$ (with $x_i \in \set{0,1}$), by using auxiliary boolean variables $y_1,\dots,y_m$, and for each $i \in \tb 1n$ a linear system $A_i \cdot y = b_i$, where $A_i$ is an $m_i \times m$ matrix over $\ZZ_2$. For the computation of $f(x_1,\dots,x_n)$, a value $x_i = 0$ means the system $A_i \cdot y = b_i$ is active, while otherwise it's inactive; the value of $f$ is $0$ if all the active systems together are unsatisfiable, and $1$ otherwise. Obviously $f$ is monotonically increasing. The task is now to put that machinery into a single system $S$ of (XOR) equations. The main idea is to ``dope'' each equation of every $A_i \cdot y = b_i$ with a dedicated new boolean variable added to the equation, making that equation trivially satisfiable, independently of everything else; all these auxiliary variables together are called $z_1, \dots, z_N$, where $N = \sum_{i=i}^n m_i$ is the number of equations in $S$.

\begin{examp}\label{exp:mspproof}
  Consider $f(x_1, x_2, x_3) = x_1 \oder x_2 \oder x_3$ ($n=3$), which can be represented by an msp with $m=2$ (with $m_1=m_2=m_3=1$, thus $N=3$), where $x_1 = 0$ activates $y_1 \oplus y_2 = 1$, while $x_2 = 0$ activates $y_1 = 0$ and $x_3 = 0$ activates $y_2 = 0$. If $x_1 = x_2 = x_3 = 0$, then the combined system is unsatisfiability, otherwise it is satisfiable. The doping process applied to $M$ yields linear equations $y_1 \oplus y_2 \oplus z_1 = 1$, $y_1 \oplus z_2 = 0$ and $y_2 \oplus z_3 = 0$.
\end{examp}

If all the doping variable used for a system $A_i \cdot y = b_i$ are set to $0$, then they disappear and the system is active, while if they are not set, then this system is trivially satisfiable, and thus is deactivated. Now consider an AC-representation $F$ of $S$. Note that the $x_i$ are not part of $F$, but the variables of $F$ are $y_1,\dots,y_m$ together with $z_1,\dots,z_N$, where the latter represent in a sense the $x_1,\dots,x_n$. From $F$ we can compute $f$ by setting the $z_j$ accordingly (if $x_i = 0$, then all $z_j$ belonging to $A_i \cdot y = b_i$ are set to $0$, if $x_i = 1$, then these variables stay unassigned), running $\rk_1$ on the system, and output $0$ iff the empty clause was produced by $\rk_1$. So we can compute msp's from AC-representations $F$ of the corresponding linear system, since we can apply partial instantiation. The second pillar of the lower-bound proof is a general polynomial-time translation of AC-representations of (arbitrary) boolean functions into monotone circuits computing a monotonisation of the boolean function (Theorem \ref{thm:acmono}; motivated by \cite{BKNW2009CircuitComplexity}), where this monotonisation precisely enables partial instantiation. So from $F$ we obtain a monotone circuit $\mc{C}$ computing $f$, whose size is polynomial in $\ell(F)$, where by \cite{BabaiGalWigderson1999MonoteSpanPrograms} the size of $\mc{C}$ is $N^{\Omega(\log N)}$ for certain msp's.

Based on \cite{GwynneKullmann2013GoodRepresentationsIII}, this superpolynomial lower bound also holds, if we consider any fixed $k \in \NNZ$, and instead of requiring unit-clause propagation to detect all forced assignments, we only ask that ``asymmetric width-bounded resolution'', i.e., $k$-resolution, is sufficient to derive all contradictions obtained by (partial) instantiation (to the variables in $S$); see Corollary \ref{cor:xorcls}. Here $k$-resolution is the appropriate generalisation of width-bounded resolution for handling long clauses (see \cite{Kl93,Ku99b,Ku00g,BeyersdorffGwynneKullmann2013PHPER}), where for each resolution step at least one parent clause has length at most $k$ (while the standard ``symmetric width'' requires both parent clauses to have length at most $k$).

\subsection{Upper bounds}
\label{sec:introupb}

Besides this fundamental negative result, we provide various forms of good representations of systems $S$ with bounded number of equations. Theorem \ref{thm:relxorcnfp} shows that there is an AC-representation with $O(n \cdot 2^m)$ many clauses. The remaining results use a stronger criterion for a ``good'' representation, namely they demand that $F \in \Propc$, where $\Propc$ is the class of ``unit-propagation complete clause-sets'' as introduced in \cite{BordeauxMarquesSilva2012KnowledgeCompilation} --- while for AC only partial assignments to the variables of $f$ are considered, now partial assignments for all variables in $F$ (which contains the variables of $f$, and possibly further auxiliary variables) are to be considered. For $m = 1$ the obvious translation $X_1$, by subdividing the big constraints into small constraints, is in $\Propc$ (Lemma \ref{lem:1softxor}). For $m = 2$ we have an intelligent representation $X_2$ in $\Propc$ (Theorem \ref{thm:2xorshared}), while the use of $X_1$ (piecewise) is still feasible for full (dag-)resolution, but not for tree-resolution (Theorem \ref{thm:2xor}).

We conjecture (Conjecture \ref{con:relxorcnfp}) that Theorem \ref{thm:relxorcnfp} and Theorem \ref{thm:2xorshared} can be combined, which would yield an fpt-algorithm for computing a representation $F \in \Propc$ for arbitrary $S$ with the parameter $m$. We now turn to a discussion of the advantages of having a representation in $\Propc$ compared to mere AC, also placing this in a wider framework.

\subsection{Measuring ``good'' representations}

We have seen yet two criteria for good representations of boolean functions, namely AC and the stronger condition of unit-propagation completeness. In Subsection \ref{sec:arcvsprop} we discuss some fundamental aspects of these two criteria, while in Subsection \ref{sec:URC} we consider another criterion, namely unit-\emph{refutation} completeness, concluding this excursion in Subsection \ref{sec:gauge} by a general framework.

\subsubsection{AC versus $\Propc$}
\label{sec:arcvsprop}

It has been shown that the practical performance of SAT solvers can depend heavily on the SAT representation used. See for example \cite{BailleuzBoufkhad2003CardinalityConstraints,Sinz2005CardinalityConstraints,Een2006Translating} for work on cardinality constraints, \cite{TamuraTagaKitagawaBanbara2009OrderEncoding,2011CompactOrderEncoding} for work on general constraint translations, and \cite{JovanovicKreuzer2010AlgAttackSAT,GwynneKullmann2011TranslationsPrelim} for investigations into different translations in cryptography. In order to obtain ``good'' representations, the basic concept is that of an AC-representation. The task is to ensure that for all (partial) assignments to the variables of the constraint, if there is a forced assignment (i.e., some variable which must be set to a particular value to avoid inconsistency), then unit-clause propagation ($\rk_1$) is sufficient to find and set this assignment. In a similar vein, there is the class $\Propc$ of propagation-complete clause-sets, containing all clause-sets for which unit-clause propagation is sufficient to detect all forced assignments; the class $\Propc$ was introduced in \cite{BordeauxMarquesSilva2012KnowledgeCompilation}, while in \cite{BBCGKV2013Propc} it is shown that membership decision is coNP-complete.

AC and $\Propc$ may at a glance seem the same concept. However there is an essential difference. When translating a constraint into SAT, typically one does not just use the variables of the constraint, but one adds auxiliary variables to allow for a compact representation. Now when speaking of AC, one only cares about assignments to the \emph{constraint variables}. But propagation-completeness deals only with the representing clause-set, thus can not know about the distinction between original and auxiliary variables, and thus it is a property on the (partial) assignments over \emph{all} variables! So a SAT representation, which is AC, will in general not fulfil the stronger property of propagation-completeness, due to assignments over both constraint \emph{and} auxiliary variables yielding a forced assignment or even an inconsistency which $\rk_1$ doesn't detect.

In \cite{JarvisaloJunttila2009LimitRestrictedLearning} it is shown that conflict-driven solvers with branching restricted to input variables have only superpolynomial run-time on $\ephp_n'$, an (extreme) Extended Resolution extension to the pigeon-hole formulas, while unrestricted branching determines unsatisfiability quickly (see \cite{BeyersdorffGwynneKullmann2013PHPER} for a proof-theoretical analysis of the general context). Also experimentally it is demonstrated in \cite{JarvisaloNiemala2008StructuralBranchingExperiments} that input-restricted branching can have a detrimental effect on solver times and proof sizes for modern CDCL solvers. This adds motivation to considering \emph{all} variables (rather than just input variables), when deciding what properties we want for SAT translations. We call this the ``absolute (representation) condition'', taking also the auxiliary variables into account, while the ``relative condition'' only considers the original variables.

Besides avoiding the creation of hard unsatisfiable sub-problems, the absolute condition also enables one to study the ``target classes'', like $\Propc$, on their own, without relation to what is represented. Target classes different from $\Propc$ have been proposed, and are reviewed in the following. The underlying idea of AC- and pro\-pa\-ga\-tion-com\-ple\-te translations is to compress all of the constraint knowledge into the SAT translation, and then to use $\rk_1$ to extract this knowledge when appropriate. In Subsection \ref{sec:URC} we present a weaker notion of what ``constraint knowledge'' could mean, while in Subsection \ref{sec:gauge} we present different extraction mechanisms.

\subsubsection{Unit-refutation completeness}
\label{sec:URC}

In \cite{GwynneKullmann2012SlurSOFSEM,GwynneKullmann2012Slur,GwynneKullmann2012SlurJ} we considered the somewhat more fundamental class $\Urefc \supset \Propc$ of ``unit-refutation complete'' clause-sets, introduced in \cite{Val1994UnitResolutionComplete} as a method for propositional knowledge compilation. Rather than requiring that $\rk_1$ detects all forced assignments (as for $\Propc$), a clause-set is in $\Urefc$ iff for all partial assignments resulting in an unsatisfiable clause-set, $\rk_1$ detects this. As shown in \cite{GwynneKullmann2012SlurSOFSEM,GwynneKullmann2012Slur,GwynneKullmann2012SlurJ}, the equation $\Urefc = \Slur$ holds, where $\Slur$, introduced in \cite{SAFS95}, is a fundamental class of clause-sets for which SAT is decidable in polynomial time; in \cite{CepekKuceraVlcek2012SLUR} it was shown that membership decision for $\Slur$ is coNP-complete.

These considerations can be extended to a general ``measurement'' approach, where we do not just have $F$ in/out for some target classes, but where a ``hardness'' measure tells us how far $F$ is from $\Propc$ resp.\ $\Urefc$ (in some sense), and this general approach is discussed next.

\subsubsection{How to gauge representations?}
\label{sec:gauge}

We now outline a a more general approach to gauge how good is a representation $F$ of a boolean function $f$. Obviously the size of $F$ must be considered, number of variables $n(F)$, number of clauses $c(F)$, number of literal occurrences $\ell(F)$. Currently we do not see a possibility to be more precise than to say that a compromise is to be sought between stronger inference properties of $F$ and the size of $F$. One criterion to judge the inference power of $F$ is AC, as already explained. This doesn't yield a possibility in case no AC-representation is feasible, nor is there a possibility for stronger representations. Our approach addresses these concerns as follows. 

\cite{GwynneKullmann2012SlurSOFSEM,GwynneKullmann2012SlurJ} introduced the measures $\hardness^V, \phardness^V, \whardness^V: \Cls \ra \NNZ$ (``hardness'', ``p-hardness'', and ``w-hardness''), where $\Cls$ is the set of all clause-sets (interpreted as CNF's), and $V$ is some set of variables. These measures determine the maximal ``effort'' (in some sense) needed to show unsatisfiability of instantiations $\vp * F$ of $F$ for partial assignments $\vp$ with $\var(\vp) \sse V$ in case of $\hardness$ and $\whardness$, resp.\ the maximal ``effort'' to determine all forced assignments for $\vp * F$ in case of $\phardness$. The ``effort'' in case of $\hardness$ or $\phardness$ is the maximal level of generalised unit-clause propagation needed, that is the maximal $k$ for reductions $\rk_k$ introduced in \cite{Ku99b,Ku00g}, where $\rk_1$ is unit-clause propagation and $\rk_2$ is (complete) elimination of failed literals. While for $\whardness$ the effort is the maximal $k$ needed for asymmetric width-bounded resolution, i.e., for each resolution step one of the parent clauses must have length at most $k$.\footnote{Symmetric width-bounded resolution requires \emph{both} parent clauses to have length at most $k$, which for arbitrary clause-length is not appropriate as complexity measure, since already unsatisfiable Horn clause-sets need unbounded symmetric width; see \cite{BeyersdorffGwynneKullmann2013PHPER} for the use of asymmetric width in the context of resolution and/or space lower bounds.}

Now we have that $F$ is an AC-representation of $f$ iff $\phardness^{\var(f)}(F) \le 1$, while $\phardness^{\var(f)}(F) \le k$ would allow higher levels of generalised unit-clause propagation (allowing potentially shorter $F$). Weaker is the requirement $\hardness^{\var(f)}(F) \le 1$, which has various names in the literature: we call it ``relative hardness'', while \cite{BordeauxJanotaMarquesSilvaMarquis2012UC} calls it ``existential unit-refutation completeness'', and \cite{Bailleux2012UnitConVsUnitProp} calls it ``unit contradiction''. Now not every forced assignment is necessarily detected by unit-clause propagation, but only unsatisfiability. Similarly, $\hardness^{\var(f)}(F) \le k$ would allow higher levels of generalised unit-clause propagation.

If we only consider ``relative (w/p-)hardness'', that is, $V = \var(f)$, then, as shown in \cite{GwynneKullmann2013GoodRepresentationsIII}, regarding polysize representations for $k \ge 1$ all conditions $\hardness^V(F) \le k$, $\phardness^V(F) \le k$, and $\whardness^V(F) \le k$ are equivalent to AC ($\phardness^V(F) \le k$), that is, the representations can be transformed in polynomial time into AC-representations. These transformations produce large representations (and very likely they are not fixed-parameter tractable in $k$), and so higher $k$ can yield smaller representations, however these savings can not be captured by the notion of polynomial size.

This situation changes, as we show in \cite{GwynneKullmann2013GoodRepresentationsI}, when we do not allow auxiliary variables, that is, we require $\var(F) = \var(f)$: Now higher $k$ for each of these measures allows short representations which otherwise require exponential size. We conjecture, that this strictness of hierarchies also holds in the presence of auxiliary variables, but using the absolute condition, i.e., $V = \var(F)$ (all variables are included in the worst-case determinations for (w/p)-hardness). The measurements in case of $V = \var(F)$ are just written as $\hardness, \phardness, \whardness: \Cls \ra \NNZ$. In this way we capture the classes $\Propc$ and $\Urefc$, namely $\Propc = \set{F \in \Cls : \phardness(F) \le 1}$ and $\Urefc = \set{F \in \Cls : \hardness(F) \le 1}$. More generally we have $\Urefc_k = \set{F \in \Cls : \hardness(F) \le k}$, $\Propc_k = \set{F \in \Cls : \phardness(F) \le k}$ and $\Wrefc_k = \set{F \in \Cls : \whardness(F) \le k}$. The basic relations between these classes are $\Wrefc_k = \Urefc_k$ for $k \le 1$, $\Propc_k \subset \Urefc_k$ for $k \ge 0$, and $\Urefc_k \subset \Wrefc_k$ for $k \ge 2$.

\subsection{Literature review}
\label{sec:litrev}

Section 1.5 of \cite{GwynneKullmann2012Slur,GwynneKullmann2012SlurJ} discusses the translation of the so-called ``Schaefer classes'' into the $\Urefc_k$ hierarchy; see Section 12.2 in \cite{DH09HBSAT} for an introduction, and see \cite{CreignouKolaitisVollmer2008ComplexityConstraints} for an in-depth overview on recent developments. All Schaefer classes except affine equations have natural translations into either $\Urefc_1$ or $\Urefc_2$. The open question was whether systems of XOR-clauses (i.e., affine equations) can be translated into $\Urefc_k$ for some fixed $k$; the current paper answers this question negatively.

Our investigations into the classes $\Urefc_k, \Propc_k, \Wrefc_k$ started with \cite{GwynneKullmann2012SlurSOFSEM,GwynneKullmann2012SlurJ}. From there on, three new developments started. First we have this paper. Then we have \cite{GwynneKullmann2013GoodRepresentationsI}, showing that without auxiliary variables, the hierarchies $\Urefc_k$, $\Propc_k$ and $\Wrefc_k$ are strict regarding polysize representations of boolean functions. Finally, \cite{GwynneKullmann2013GoodRepresentationsIII} discusses general tools for obtaining ``good'' representations, and how SAT solvers perform with them; it contains the proof that regarding the relative condition and allowing auxiliary variables, all three hierarchies $\Urefc_k$, $\Propc_k$ and $\Wrefc_k$ collapse to their first level (regarding polysize representations). The predecessor of \cite{GwynneKullmann2013GoodRepresentationsI,GwynneKullmann2013GoodRepresentationsIII} and the current report is \cite{GwynneKullmann2013GoodRepresentations}, while the (shortened) conference version of the current report is \cite{GwynneKullmann2013GoodRepresentationsIILata}.

\subsubsection{Discussion of terminology ``arc-consistent representation''}
\label{sec:introAC}

We think that the current terminology in the literature can potentially cause confusion between the fields CSP (constraint-satisfaction problems) and SAT, and thus some clarifying discussion is needed. The notion ``AC-representation'' $F$ of $f$ fully expanded says: ``a representation maintaining hyperarc-consistency (or generalised arc-consistency) via unit-clause propagation for the (single, global) boolean constraint $f$ after (arbitrary) partial assignments''. Recall that a constraint is ``hyperarc-consistent'' if for each variable each value in its (current) domain is still available (does not yield an inconsistency); see Chapter 3 of \cite{RBW2006HandbookCSP}. However for a (boolean) clause-set $F$ it is not clear what the ``constraint'' is. In our context it is most natural to consider the whole $F$ as a ``constraint'', and then, since every variable has (precisely) two values, ``arc-consistency'' of $F$ means that $F$ has no forced assignments (or no ``forced literals''). It must be emphasised here that considering $F$ as a single constraint is not a natural point of view for the CSP area. The standard notion of ``arc-consistency'' just applies to $n(F)=2$, and for ordinary constraints $n(F)$ is considered as constant --- only ``global constraints'' are allowed to contain a non-constant number of variables. Especially for XOR-clause-sets it is tempting to take each XOR-clause as a constraint, but this is not interesting here.

The notion of ``an encoding maintaining arc-consistency via unit propagation'' has been introduced in \cite{Gent2002ArcConsistency}, showing that the support encoding of a single constraint yields in our terminology an AC-representation --- it is essential here that this assumes as usual that the number of variables is constant. ``Maintaining'', as in ``MAC'' for ``maintaining arc-consistency'', applies to constraints after a domain restriction (which for SAT is achieved by partial assignments), where (hyper)arc-consistency has to be re-established (this can be done in polynomial time, since the number of variables in a constraint is constant). Apparently the first explicit definition of ``arc-consistency under unit propagation'' for SAT representations is \cite{Een2006Translating}, the Definition on Page 5 (it is left open whether $\sigma$ may also involve the introduced variables, but this is a kind of automatic assumption, since only the variables of the (original) constraint are considered in this context\footnote{An assumption which we challenge by considering $\Propc$.}). For further examples for pseudo-boolean constraints see Section 22.6.7 in \cite{RM09HBSAT} and \cite{BailleuzBoufkhad2003CardinalityConstraints,Sinz2005CardinalityConstraints,BailleuxBoufkhadRoussel2009PBCNF}, while related considerations one finds in \cite{BarahomaJungKatsirelosWalsh2008EncodingDNNF}.

We prefer to speak of ``AC-representations'', hiding the ``arc-consistency''. It also seems superfluous to mention in this context ``unit(-clause) propagation''. One could also say ``AC-translation'' or``AC-encoding'', but we reserve ``translation'' for (poly-time) functions computing a representation, and ``encoding'' for the translation of non-boolean variables into boolean variables. Our own ``proper terminology'' for an ``AC-representation'' $F$ of $f$ is that $F$ is a CNF-representation of $f$ with relative p-hardness at most $1$, i.e., $\phardness^{\var(f)}(F) \le 1$.

\subsubsection{Applications of XOR-constraints}

If we do not specify the representation in the following, then essentially what we call $X_1$ is used, that is, breaking up long XOR-constraints into short ones and using the unique equivalent CNF's for the short constraints.

XOR-constraints are a typical part of cryptographic schemes, and accordingly it is important to have ``good'' representations for them. The earliest application of SAT to cryptanalysis is \cite{MassacciMarraro2000DESSAT}, translating DES to SAT and then considering finding a key. In \cite{BardCourtois2007AlgebraicDES}, DES is encoded to ANF (``algebraic normal form'', that is, XOR's of conjunctions), and then translated. \cite{JovanovicKreuzer2010AlgAttackSAT} attacks DES, AES and the Courtois Toy Cipher via translation to SAT. Each cipher is  first translated to equations over GF(2) and then to CNF. A key contribution is a specialised translation of certain forms of polynomials, designed to reduce the number of variables and clauses. The size for breaking up long XOR-constraints is called the ``cutting length'', and has apparently some effect on solver times. \cite{MironovZhang2006SATHash} translates MD5 to SAT and finds collisions. MD5 is translated by modelling it as a circuit (including XORs) and applying the Tseitin translation.

\cite{MarquesSilvaSakallah2000EDA} provides an overview of SAT-based methods in Electronic Design Automation, and suggests keeping track of circuit information (fan in/fan out of gates etc.) in the SAT solver when solving such instances. XOR is relevant here due to the use of XOR gates in the underlying circuit being checked (and translated).

A potential application area is the translation of pseudo-boolean constraints, as investigated by \cite{Een2006Translating}. Translations via ``full-adders'' introduces XORs via translation of the full-adder circuit. It is shown that this translation does not produce an AC-representation (does not ``maintain arc-consistency via unit propagation''), and the presence of XOR and the log encoding is blamed for this (in Section 5.5). Experiments conclude that sorting network and BDD methods perform better, as long as their translations are not too large.

\subsubsection{Hard examples via XORs}

It is well-known that translating each XOR to its prime implicates results in hard instances for resolution. This goes back to the ``Tseitin formulas'' introduced in \cite{Ts68}, which were proven hard for full resolution in \cite{Urq87}, and generalised to (empirically) hard satisfiable instances in \cite{HaanpaaJarvisaloKaskiNiemela2006HardSATBench}. A well-known benchmark was introduced in \cite{CrawfordKearns1994MinimalParityDisagreement}, called the ``Minimal Disagreement Parity'' problem (which became the \texttt{parity32} benchmarks in the SAT2002 competition).  Given $m$ vectors $\vec{x}_i \in \set{0,1}^n$, further $m$ bits $y_i \in \set{0,1}$, and $k \in \NNZ$, find a vector $\vec{a} \in \set{0,1}^n$ such that $\abs{\set{i : \vec{a} \cdot \vec{x}_i \not= y_i}} \le k$, where $\vec{a} \cdot \vec{x}_i$ is the scalar product. The SAT encoding is $X_1(r_i \oplus y_i = (\vec{a}_1 \wedge (\vec{x}_i)_1) \oplus \dots \oplus (\vec{a}_n \wedge (\vec{x}_i)_n))$ for $i = 1, \dots, m$ (so $r_i = 0$ iff $y_i = \vec{a} \cdot \vec{x}_i$), together with a cardinality constraint $\sum_{1 \le i \le m} r_i \le k$ based on full-adders. So the XORs occur both in the summations and the cardinality constraint. These benchmarks were first solved by the solver \texttt{EqSatz} (\cite{Li2000Equivalency}).

\subsubsection{Special reasoning}
\label{sec:introspecreas}

It is natural to consider extensions of resolution and/or SAT techniques to handle XOR-constraints more directly. The earliest theoretical approach seems \cite{BaumgartnerMassacci2000TamingXOR}, integrating a proof calculus for Gaussian elimination with an abstract proof calculus modelling DPLL. It is argued that such a system should offer improvements over just DPLL/resolution in handling XORs. \cite{Heule2004Diplom} points out the simple algorithm for extracting ``equivalence constraints''. The earliest SAT solver with special reasoning is \texttt{EqSatz} (\cite{Li2000Equivalency}), extracting XOR-clauses from its input and applying DP-resolution plus incomplete XOR reasoning rules. More recently, \texttt{CryptoMiniSAT} (\cite{CryptoSAT2009,Soos2010SATGauss}) integrates Gaussian elimination during search, allowing both explicitly specified XOR-clauses and also XOR-clauses extracted from CNF input. However in the newest version 3.3 the XOR handling during search is removed, since it is deemed too expensive.\footnote{See \url{http://www.msoos.org/2013/08/why-cryptominisat-3-3-doesnt-have-xors/}.} Further approaches for hybrid solvers one finds in \cite{Chen2009HybridXOR,HanJiang2012GaussianSAT}.

A systematic study of the integration of XOR-reasoning and SAT-techniques has been started with \cite{Laitinen2010DPLLParity}, by introducing the ``DPLL(XOR)'' framework, similar to SMT. These techniques have also been integrated into \texttt{MiniSat}. \cite{LaitinenJunttilaNiemela2011EqParReasDPLLXOR} expands on this by reasoning about equivalence classes of literals created by binary XORs, while \cite{LaitinenJunttilaNiemela2012ConflictXORLearn} learns conflicts in terms of ``parity (XOR) explanations''. The latest paper \cite{Laitinen2012DPLLParity} (with underlying report \cite{Laitinen2012DPLLParityE}) extends the reasoning from ``Gau\ss{} elimination'' to ``Gau\ss{}-Jordan elimination'', which corresponds to moving from relative hardness to relative p-hardness, i.e., also detecting forced literals, not just inconsistency.\footnote{We say ``relative'' here, since the reasoning mechanism is placed outside of SAT solving, different from the ``absolute'' condition, where also the reasoning itself is made accessible to SAT solving (that is, one can (feasibly!) split in some sense on the higher-level reasoning).} Theorem 4 in \cite{Laitinen2012DPLLParity} is similar in spirit to Corollary \ref{cor:1acylcprop}, Part \ref{cor:1acylcprop2}, considering conditions when strong reasoning only needs to be applied to ``components''.

Altogether we see a mixed picture regarding special reasoning in SAT solvers. The first phase of expanding SAT solvers could be seen as having ended in some disappointment regarding XOR reasoning, but with \cite{Laitinen2010DPLLParity} a systematic approach towards integrating special reasoning has been re-opened. A second approach for handling XOR-constraints, the approach of this paper, is by using intelligent translations (possibly combined with special reasoning).

\subsubsection{Translations to CNF}

Switching now to translations of XORs to CNF, \cite{LaitinenJunttilaNiemelae2012Parity} identifies the subsets of ``tree-like'' systems of XOR constraints, where the standard translation delivers an AC-representation (our Theorem \ref{thm:suffx1pc} strengthens this, showing that indeed a representation in $\Propc$ is obtained):
\begin{itemize}
\item \cite{LaitinenJunttilaNiemelae2012Parity} also considered equivalence reasoning, where for ``cycle-partitionable'' systems of XOR constraints this reasoning suffices to derive all conclusions.
\item Furthermore \cite{LaitinenJunttilaNiemelae2012Parity} showed how to eliminate the need for such special equivalence reasoning by another AC-representation.
\item In general, the idea is to only use Gaussian elimination for such parts of XOR systems which the SAT solver is otherwise incapable of propagating on. Existing propagation mechanisms, especially unit-clause propagation, and to a lesser degree equivalence reasoning, are very fast, while Gaussian elimination is much slower (although still poly-time).
\end{itemize}
Experimental evaluation on SAT 2005 benchmarks instances showed that, when ``not too large'', such CNF translations outperform dedicated XOR reasoning modules. The successor \cite{LaitinenJunttilaNiemelae2013Parity} provides several comparisons of special-reasoning machinery with resolution-based methods, and in Theorem 4 there we find a general AC-translation; our Theorem \ref{thm:relxorcnfp} yields a better upper bound, but the heuristic reasoning of \cite{Laitinen2012DPLLParity,LaitinenJunttilaNiemelae2013Parity} seems valuable, and should be explored further.

\subsection{Better understanding of ``SAT''}
\label{sec:introbs}

One motivation of our investigations is the question on the relation between CSP and SAT, and on the success of SAT. Viewing a linear system $S$ as a constraint on $\var(S)$, one can encode evaluation via Tseitin's translation, obtaining a CNF-representation $F$ with the property that for every \emph{total} assignment $\vp$, i.e., $\var(\vp) = \var(S)$, we have that $\rk_1(\vp * F)$ either contains the empty clause or is empty.\footnote{In Subsection 9.4.1 of \cite{GwynneKullmann2013GoodRepresentationsI} this class of representations is called $\ex\mc{UP}$; up to linear-time transformation it is the same as representations by boolean circuits.} However this says nothing about \emph{partial} assignments $\vp$, and as our result shows (Theorem \ref{thm:xorclsrel}), there is indeed no polysize representation which handles all partial assignments. One can write an algorithm (a kind of ``constraint'', now using Gaussian elimination) which handles (in a sense) all partial assignments in polynomial time (detects unsatisfiability of $\vp * F$ for all partial assignments $\vp$), but this can not be integrated into the CNF formalism (by using auxiliary variables and clauses), since algorithms always need total assignments, and so partial assignments $\vp$ would need to be encoded --- the information ``variable $v$ not assigned'' (i.e., $v \notin \var(\vp)$) needs to be represented by \emph{setting} some auxiliary variable, and this must happen by a mechanism outside of the CNF formalism.

It is an essential strength of the CNF formalism to allow partial instantiation; if we want these partial instantiations also to be easily understandable by a SAT solver, then the results of \cite{BKNW2009CircuitComplexity} and our results show that there are restrictions. Yet there is little understanding of these restrictions. There are many examples where AC and stronger representations are possible, while the current non-representability results, one in \cite{BKNW2009CircuitComplexity}, one in this article and a variation on \cite{BKNW2009CircuitComplexity} in \cite{BeyersdorffGwynneKullmann2013PHPER}, rely on non-trivial lower bounds on monotone circuit complexity; in fact Corollary \ref{cor:otherdir} shows that there is a polysize AC-representation of a boolean function $f$ if and only if the monotonisation $\widehat{f}$, which encodes partial assignments to $f$, has polysize monotone circuits.

\subsection{Overview on results}

In Section \ref{sec:prelim} we give the basic definitions related to clause-sets, and in Section \ref{sec:measurerepcomp} we review the basic concepts and notions related to hardness, w-hardness, and the classes $\Urefc_k$ and $\Wrefc_k$. Section \ref{sec:propc} contains a review of p-hardness and the classes $\Propc_k$, and also introduces basic criteria for $\bc_{i \in I} F_i \in \Propc_k$ for clause-sets $F_i \in \Propc_k$: the main result Theorem \ref{thm:acylcprop} shows that the ``incidence graph'' being acyclic is sufficient. In Section \ref{sec:xorclausesets} we introduce the central concepts of this article, ``XOR-clause-sets'' and their CNF-representations, and show in Lemma \ref{lem:characimplxor}, that the sum of XOR-clauses is the (easier) counterpart to the resolution operation for (ordinary) clauses.

In Section \ref{sec:noarccons} we present our general lower bound. Motivated by \cite{BKNW2009CircuitComplexity}, in Theorem \ref{thm:acmono} we show that from a CNF-representation of relative hardness $1$ of a boolean function $f$ we obtain in polynomial time a monotone circuit computing the monotonisation $\widehat{f}$, which extends $f$ by allowing partial assignments to the inputs.\footnote{The precise relation to the results of \cite{BKNW2009CircuitComplexity} is not clear. The notion of ``CNF decomposition of a consistency checker'' in \cite{BKNW2009CircuitComplexity} is similar to an AC-representation, but it contains an additional special condition.} Actually, as we show in Corollary \ref{cor:otherdir}, in this way AC-representations are equivalently characterised. Theorem \ref{thm:xorclsrel} shows that there are no short AC-representations of arbitrary XOR-clause-sets $F$ (at all), with Corollary \ref{cor:xorcls} generalising this to arbitrary relative w-hardness.

The fundamental translation $X_0$ of XOR-clause-sets (using the trivial translation for every XOR-clause) is studied in Section \ref{sec:transx0}, with Lemma \ref{lem:suffx0pc} stating the basic criterion for $X_0(F) \in \Propc$. Furthermore the Tseitin formulas are discussed. The standard translation, called $X_1$, uses $X_0$, but breaks up long clauses first (to avoid the exponential size-explosion), and is studied in Section \ref{sec:transx1}. The main result here is Theorem \ref{thm:suffx1pc}, showing that if $F$ fulfils the basic graph-theoretic properties considered before, then $X_1(F) \in \Propc$. In Section \ref{sec:transarbxor} we show that $X_1(F^*)$, where $F^*$ is obtained from $F$ by adding all implied XOR-clauses, achieves AC in linear time for a constant number of XOR-clauses (Theorem \ref{thm:relxorcnfp}). In Section \ref{sec:transtxor} we turn to the question of two XOR-clauses $C, D$ and $F = \set{C,D}$. In Theorem \ref{thm:2xorshared} we show how to obtain a translation $X_2(C,D) \in \Propc$. Then we discuss $X_1(F)$ and $X_1(F^*)$ and show, that all three cases can be distinguished here regarding their complexity measures; the worst representation is $X_1(F)$, which still yields an acceptable translation regarding w-hardness, but not regarding hardness (Theorem \ref{thm:2xor}).  Finally in Section \ref{sec:open} we present the conclusions and open problems.

\section{Preliminaries}
\label{sec:prelim}

We follow the general notations and definitions as outlined in \cite{Kullmann2007HandbuchMU}. We use $\NN = \set{1,2,\dots}$ and $\NNZ = \NN \cup \set{0}$. Let $\Va$ be the infinite set of variables, and let $\Lit = \Va \cup \set{\ol{v} : v \in \Va}$ be the set of literals, the disjoint union of variables as positive literals and complemented variables as negative literals. We use $\ol{L} := \set{\ol{x} : x \in L}$ to complement a set $L$ of literals. A clause is a finite subset $C \subset \Lit$ which is complement-free, i.e., $C \cap \ol{C} = \es$; the set of all clauses is denoted by $\Cl$. A clause-set is a finite set of clauses, the set of all clause-sets is $\Cls$. By $\var(x) \in \Va$ we denote the underlying variable of a literal $x \in \Lit$, and we extend this via $\var(C) := \set{\var(x) : x \in C} \subset \Va$ for clauses $C$, and via $\var(F) := \bc_{C \in F} \var(C)$ for clause-sets $F$. The possible literals in a clause-set $F$ are denoted by $\lit(F) := \var(F) \cup \ol{\var(F)}$. Measuring clause-sets happens by $n(F) := \abs{\var(F)}$ for the number of variables, $c(F) := \abs{F}$ for the number of clauses, and $\ell(F) := \sum_{C \in F} \abs{C}$ for the number of literal occurrences. A special clause-set is $\top := \es \in \Cls$, the empty clause-set, and a special clause is $\bot := \es \in \Cl$, the empty clause.

A partial assignment is a map $\vp: V \ra \set{0,1}$ for some finite $V \subset \Va$, where we set $\var(\vp) := V$, and where the set of all partial assignments is $\Pass$. For $v \in \var(\vp)$ let $\vp(\ol{v}) := \ol{\vp(v)}$ (with $\ol{0} = 1$ and $\ol{1} = 0$). We construct partial assignments by terms $\pab{x_1 \ra \ve_1, \dots, x_n \ra \ve_n} \in \Pass$ for literals $x_1, \dots, x_n$ with different underlying variables and $\ve_i \in \set{0,1}$. For $\vp \in \Pass$ and $F \in \Cls$ we denote the result of applying $\vp$ to $F$ by $\vp * F$, removing clauses $C \in F$ containing $x \in C$ with $\vp(x) = 1$, and removing literals $x$ with $\vp(x) = 0$ from the remaining clauses. By $\Sat := \set{F \in \Cls \mb \ex\, \vp \in \Pass : \vp * F = \top}$ the set of satisfiable clause-sets is denoted, and by $\Usat := \Cls \sm \Sat$ the set of unsatisfiable clause-sets.

 By $\bmm{\rk_1}: \Cls \ra \Cls$ we denote unit-clause propagation, that is,
 \begin{itemize}
 \item $\rk_1(F) := \set{\bot}$ if $\bot \in F$,
 \item $\rk_1(F) := F$ if $F$ contains only clauses of length at least $2$,
 \item while otherwise a unit-clause $\set{x} \in F$ is chosen, and recursively we define $\rk_1(F) := \rk_1(\pao x1 * F)$.
 \end{itemize}
 It is easy to see that the final result $\rk_1(F)$ does not depend on the choices of the unit-clauses. In \cite{Ku99b,Ku00g} the theory of generalised unit-clause propagation $\bmm{\rk_k}: \Cls \ra \Cls$ for $k \in \NNZ$ was developed; the basic idea should become clear from the definition of $\rk_2(F)$, which is complete ``failed literal elimination'': if $\rk_1(F) = \set{\bot}$, then $\rk_2(F) := \set{\bot}$, if otherwise there is a literal $x \in \lit(F)$ such that $\rk_1(\pao x0 * F) = \set{\bot}$, then $\rk_2(F) := \rk_2(\pao x1 * F)$, and otherwise $\rk_2(F) := F$.

Reduction by $\rk_k$ applies certain \textbf{forced assignments} to the (current) $F$, which are assignments $\pao x1$ such that the opposite assignment yields an unsatisfiable clause-set, that is, where $\pao x0 * F \in \Usat$; the literal $x$ here is also called a \textbf{forced literal}. The reduction applying all forced assignments is denoted by $\rki: \Cls \ra \Cls$ (so $F \in \Usat \Lra \rki(F) = \set{\bot}$). Forced assignments are also known under other names, for example ``necessary assignments'' or ``backbones''; see \cite{JanotaLycneMarquesSilva2012Backbones} for an overview on algorithms computing all forced assignments.

Two clauses $C, D \in \Cl$ are resolvable iff they clash in exactly one literal $x$, that is, $C \cap \ol{D} = \set{x}$, in which case their resolvent is $\bmm{C \res D} := (C \cup D) \sm \set{x,\ol{x}}$ (with resolution literal $x$). A resolution tree is a full binary tree formed by the resolution operation. We write \bmm{T : F \vdash C} if $T$ is a resolution tree with axioms (the clauses at the leaves) all in $F$ and with derived clause (at the root) $C$.

A \emph{prime implicate} of $F \in \Cls$ is a clause $C$ such that a resolution tree $T$ with $T: F \vdash C$ exists, but no $T'$ exists for some $C' \subset C$ with $T': F \vdash C'$; the set of all prime implicates of $F$ is denoted by $\bmm{\primec_0(F)} \in \Cls$. The term ``implicate'' refers to the implicit interpretation of $F$ as a conjunctive normal form (CNF). Considering clauses as combinatorial objects one can speak of ``prime clauses'', and the ``$0$'' in our notation reminds of ``unsatisfiability'', which is characteristic for CNF. Two clause-sets $F, F' \in \Cls$ are equivalent iff $\primec_0(F) = \primec_0(F')$. A clause-set $F$ is unsatisfiable iff $\primec_0(F) = \set{\bot}$. If $F$ is unsatisfiable, then every literal $x \in \Lit$ is a forced literal for $F$, while otherwise $x$ is forced for $F$ iff $\set{x} \in \primec_0(F)$. It can be considered as known, that for a clause-set $F$ the computation of all prime implicates is fixed-parameter tractable (fpt) in the number $c(F)$ of clauses, but perhaps that is not stated explicitly in the literature, and so we provide a simple proof:
\begin{lem}\label{lem:primcec}
  Consider $F \in \Cls$ and let $F' := \primec_0(F)$. Then $c(F') \le 2^{c(F)}-1$, and $F'$ can be computed in time $O(\ell(F) \cdot 2^{3c(F)})$.
\end{lem}
\begin{prf}
By Subsection 4.1 (especially Lemma 4.12) in \cite{GwynneKullmann2013GoodRepresentationsI} we run through all subsets $G \sse F$, determine the set $C$ of pure literals of $G$, and include $C$ if $C$ is an implicate of $F$ (note that SAT-decision for a CNF-clause-set $F \in \Cls$ can be done in time $O(\ell(F) \cdot 2^{c(F)})$). To the final result subsumption-elimination is applied, which can be done in cubic time, and we obtain $\primec_0(F)$. \Qed
\end{prf}

\section{Measuring ``SAT representation complexity''}
\label{sec:measurerepcomp}

In this section we define and discuss the measures $\hardness, \whardness: \Cls \ra \NNZ$ and the corresponding classes $\Urefc_k \sse \Wrefc_k \subset \Cls$. It is mostly of an expository nature, explaining the background from \cite{Ku99b,Ku00g,GwynneKullmann2012SlurSOFSEM,GwynneKullmann2012Slur,GwynneKullmann2012SlurJ}. For the measure $\phardness: \Cls \ra \NNZ$ and the corresponding classes $\Propc_k$ see Section \ref{sec:propc}.

\subsection{Hardness and $\Urefc_k$}
\label{sec:prelimhdUC}

Hardness for unsatisfiable clause-sets was introduced in \cite{Ku99b,Ku00g}, while the specific generalisation to arbitrary clause-sets used here was first mentioned in \cite{AnsoteguiBonetLevyManya2008Hardness}, and systematically studied in \cite{GwynneKullmann2012SlurSOFSEM,GwynneKullmann2012Slur,GwynneKullmann2012SlurJ}. Using the Horton-Strahler number $\hts(T)$ of binary trees, applied to resolution trees $T: F \vdash C$ (deriving clause $C$ from $F$), the hardness $\hardness(F)$ for $F \in \Cls$ can be defined as the minimal $k \in \NNZ$ such that for all prime implicates $C$ of $F$ there exists $T : F \vdash C$ with $\hts(T) \le k$. An equivalent characterisation uses necessary levels of generalised unit-clause propagation (see \cite{GwynneKullmann2012SlurSOFSEM,GwynneKullmann2012Slur,GwynneKullmann2012SlurJ} for the details):
\begin{defi}\label{def:charachd}
  Consider the reductions $\rk_k: \Cls \ra \Cls$ for $k \in \NNZ$ as introduced in \cite{Ku99b}; it is $\rk_1$ unit-clause propagation, while $\rk_2$ is (full, iterated) failed-literal elimination. Then $\bmm{\hardness^V(F)} \in \NNZ$ for $F \in \Cls$ and $V \sse \Va$ is the minimal $k \in \NNZ$ such that for all $\vp \in \Pass$ with $\var(\vp) \sse V$ and $\vp * F \in \Usat$ holds $\rk_k(\vp * F) = \set{\bot}$, i.e., the minimal $k$ such that $\rk_k$ detects unsatisfiability of any partial instantiation of variables in $V$. Furthermore $\bmm{\hardness(F)} := \hardness^{\var(F)}(F)$.
\end{defi}
For every $F \in \Cls$ there is a partial assignment $\vp$ with $\vp * F = \rk_k(F)$, where $\vp$ consists of certain forced assignments $\pao x1 \sse \vp$, i.e., $\pao x0 * F \in \Usat$. A weaker localisation of forced assignments has been considered in \cite{HaanKanjSzeider2013BackbonesC}, namely ``$k$-backbones'', which are forced assignments $\pao x1$ for $F$ such that there is $F' \sse F$ with $c(F') \le k$ and such that $\pao x1$ is forced also for $F'$. It is not hard to see that $\rk_k$ for $k \in \NNZ$ will set all $k$-backbones of $F \in \Cls$ (using that for $F \in \Usat$ we have $\hardness(F) < c(F)$ by Lemma 3.18 in \cite{Ku99b}). The fundamental level of ``hardness'' for forced assignments or unsatisfiability is given by the level $k$ needed for $\rk_k$. As a hierarchy of CNF-classes and only considering detection of unsatisfiability (for all instantiations), this is captured by the $\Urefc_k$-hierarchy (with ``UC'' for ``unit-refutation complete''):
\begin{defi}\label{def:UC}
  For $k \in \NNZ$ let $\bmm{\Urefc_k} := \set{F \in \Cls : \hardness(F) \le k}$.
\end{defi}
$\Urefc_1 = \Urefc$ is the class of unit-refutation complete clause-sets, as introduced in \cite{Val1994UnitResolutionComplete}. In \cite{GwynneKullmann2012SlurSOFSEM,GwynneKullmann2012Slur,GwynneKullmann2012SlurJ} we show that $\Urefc = \Slur$, where $\Slur$ is the class of clause-sets solvable via Single Lookahead Unit Resolution (see \cite{FrGe98}). Using \cite{CepekKuceraVlcek2012SLUR} we then obtain (\cite{GwynneKullmann2012SlurSOFSEM,GwynneKullmann2012Slur,GwynneKullmann2012SlurJ}) that membership decision for $\Urefc_k$ ($ = \Slur_k$) is coNP-complete for $k \ge 1$. The class $\Urefc_2$ is the class of all clause-sets where unsatisfiability for any partial assignment is detected by failed-literal reduction (see Section 5.2.1 in \cite{HvM09HBSAT} for the usage of failed literals in SAT solvers).

\subsection{W-Hardness and $\Wrefc_k$}
\label{sec:prelimwhdWC}

A basic weakness of the standard notion of width-restricted resolution, which demands that \emph{both} parent clauses must have length at most $k$ for some fixed $k \in \NNZ$ (``width'', denoted by $\wid(F)$ below; see \cite{SW98}), is that even Horn clause-sets require unbounded width in this sense. A better solution, as investigated and discussed in \cite{Ku99b,Ku00g,BeyersdorffGwynneKullmann2013PHPER}, seems to use the notion of ``$k$-resolution'' as introduced in \cite{Kl93}, where only \emph{one} parent clause needs to have length at most $k$ (thus properly generalising unit-resolution).\footnote{Symmetric width is only applied to clause-sets with bounded clause-length, and here everything can be done as well via asymmetric width, as discussed in \cite{BeyersdorffGwynneKullmann2013PHPER}. It might be that symmetric width could have a relevant combinatorial meaning, so that symmetric and asymmetric width both have their roles.} Nested input-resolution (\cite{Ku99b,Ku00g}) is the proof-theoretic basis of hardness, and approximates tree-resolution. In the same vein, $k$-resolution is the proof-theoretic basis of ``w-hardness'', and approximates dag-resolution (see Theorem 6.12 in \cite{Ku00g}):
\begin{defi}\label{def:whd}
  The \textbf{w-hardness} $\bmm{\whardness}: \Cls \ra \NNZ$ (``width-hardness'', or ``asymmetric width'') is defined for $F \in \Cls$ as follows:
  \begin{enumerate}
  \item If $F \in \Usat$, then $\whardness(F)$ is the minimum $k \in \NNZ$ such that $k$-resolution refutes $F$, that is, such that $T : F \vdash \bot$ exists where for each resolution step $R = C \res D$ in $T$ we have $\abs{C} \le k$ or $\abs{D} \le k$ (this concept corresponds to Definition 8.2 in \cite{Ku99b}, and is a special case of ``$\mr{wid}_{\mc{U}}$'' as introduced in Subsection 6.1 of \cite{Ku00g}).
  \item If $F = \top$, then $\whardness(F) := 0$.
  \item If $F \in \Sat \sm \set{\top}$, then $\DST \whardness(F) := \max_{\vp \in \Pass} \set{\whardness(\vp * F) : \vp * F \in \Usat}$.
  \end{enumerate}
  For $k \in \NNZ$ let $\bmm{\Wrefc_k} := \set{F \in \Cls : \whardness(F) \le k}$.

  The \textbf{symmetric width} $\bmm{\wid}: \Cls \ra \NNZ$ is defined in the same way, only that for $F \in \Usat$ we define $\wid(F)$ as the minimal $k \in \NNZ$ such that there is $T : F \vdash \bot$, where all clauses of $T$ (axioms and resolvents) have length at most $k$.

  More generally, for $V \sse \Va$, we define $\bmm{\whardness^V(F)} := \whardness(F)$ and $\bmm{\wid^V(F)} := \wid(F)$ for unsatisfiable $F$, while for satisfiable $F$ only $\vp \in \Pass$ with $\var(\vp) \sse V$ are considered.
\end{defi}
We have $\Wrefc_0 = \Urefc_0$, $\Wrefc_1 = \Urefc_1$, and for all $k \in \NNZ$ holds $\Urefc_k \sse \Wrefc_k$ (this follows by Lemma 6.8 in \cite{Ku00g} for unsatisfiable clause-sets, which extends to satisfiable clause-sets by definition). What is the relation between asymmetric width $\whardness(F)$ and the well-known (for unsatisfiable $F$) symmetric width $\wid(F)$ ? By definition we have $\whardness(F) \le \wid(F)$ for all $F \in \Cls$. Now consider $F \in \Ho \cap \Usat$, where $\Ho$ is the set of all Horn clause-sets, defined by the condition that each clause contains at most one positive literal. The symmetric width $\wid(F)$ here is unbounded, and is equal to the maximal clause-length of $F$ in case $F$ is minimally unsatisfiable. But we have $\whardness(F) \le 1$ here. So for unbounded clause-length there is an essential difference between symmetric and asymmetric width. On the other hand we have
\begin{displaymath}
  \wid(F) \le \whardness(F) + \max(\whardness(F),p)
\end{displaymath}
for $F \in \Pcls{p} := \set{F \in \Cls \mb \fa\, C \in F : \abs{C} \le p}$, $p \in \NNZ$, by Lemma 8.5 in \cite{Ku99b}, or, more generally, Lemma 6.22 in \cite{Ku00g} (also shown in \cite{BeyersdorffGwynneKullmann2013PHPER}). So for bounded clause-length and considered asymptotically, symmetric and asymmetric width can be considered equivalent.

\section{P-Hardness and $\Propc_k$}
\label{sec:propc}

Complementary to ``unit-refutation completeness'', there is the notion of ``pro\-pa\-ga\-tion-com\-ple\-te\-ness'' as investigated in \cite{DarwichePipatsrisawat2011ClauseLearnRes,BordeauxMarquesSilva2012KnowledgeCompilation}, yielding the class $\Propc \subset \Urefc$. This was captured and generalised by a measure $\phardness: \Cls \ra \NNZ$ of ``propagation-hardness'' along with the associated hierarchy, defined in \cite{GwynneKullmann2012Slur,GwynneKullmann2012SlurJ} as follows:
\begin{defi}\label{def:phardness}
  For $F \in \Cls$ and $V \sse \Va$ we define the (relative) \textbf{propagation-hardness} (for short ``p-hardness'') $\bmm{\phardness^V(F)} \in \NNZ$ as the minimal $k \in \NNZ$ such that for all partial assignments $\vp \in \Pass$ with $\var(\vp) \sse V$ we have $\rk_k(\vp * F) = \rki(\vp * F)$, where $\rki: \Cls \ra \Cls$ applies all forced assignments, and can be defined by $\rki(F) := \rk_{n(F)}(F)$. Furthermore $\bmm{\phardness(F)} := \phardness^{\var(F)}(F)$. For $k \in \NNZ$ let $\bmm{\Propc_k} := \set{F \in \Cls : \phardness(F) \le k}$ (the class of \textbf{propagation-complete clause-sets of level $k$}).
\end{defi}
Remarks:
\begin{enumerate}
\item We have $\Propc = \Propc_1$.
\item For $k \in \NNZ$ we have $\Propc_k \subset \Urefc_k \subset \Propc_{k+1}$.
\item By definition (and composition of partial assignments) we have that all classes $\Propc_k$ are stable under application of partial assignments.
\item We have $F \in \Propc_k$ iff for all $\vp \in \Pass$ the clause-set $F' := \rk_k(\vp * F)$ in case of $F' \ne \set{\bot}$ has no forced literals.
\end{enumerate}

Recall that a clause-set $F$ has no forced assignments (at all) if and only if all prime implicates of $F$ have length at least $2$.\footnote{The ``at all'' is for the case $F = \set{\bot}$, where every literal is forced for $F$, but $F$ has no literals.} Before proving the main lemma (Lemma \ref{lem:acyclpropc}), we need a simple characterisation of such clause-sets without forced assignments. Recall that a partial assignment $\vp$ is an autarky for $F \in \Cls$ iff for all $C \in F$ with $\var(\vp) \cap \var(C) \ne \es$ holds $\vp * \set{C} = \top$; for an autarky $\vp$ for $F$ the (sub-)clause-set $\vp * F$ is satisfiable iff $F$ is satisfiable. See \cite{Kullmann2007HandbuchMU} for the general theory of autarkies.
\begin{lem}\label{lem:characnofa}
  A clause-set $F \in \Cls$ has no forced assignments if and only if $F$ is satisfiable, and for every $x \in \lit(F)$ there is an autarky $\vp$ for $F$ with $\vp(x)=1$.
\end{lem}
\begin{prf}
If $F$ has no forced assignment, then $F$ can not be unsatisfiable (since then every literal would be forced), and for a literal $x \in \lit(F)$ the clause-set $\pao x1 * F$ is satisfiable (since $\ol{x}$ is not forced), thus has a satisfying assignment $\vp$, and the composition $\vp \circ \pao x1$ (first assigning $x$, then applying $\vp$) is an autarky for $F$. Now let $F$ be satisfiable, and for each $x \in \lit(F)$ let there be an autarky $\vp$ for $F$ with $\vp(x)=1$. If $F$ had a forced literal $x$, then consider an autarky $\vp$ for $F$ with $\vp(\ol{x}) = 1$: since $x$ is forced, $\pao x0 * F$ is unsatisfiable, while by the autarky condition $\vp * F$ would be satisfiable. \Qed
\end{prf}

In the rest of this section we show that having an ``acyclic incidence graph'' yields a sufficient criterion for $\bc_{i \in I} F_i \in \Propc_k$ for clause-sets $F_i \in \Propc_k$.

\begin{defi}\label{def:incgr}
  For a finite family $(F_i)_{i \in I}$ of clause-sets $F_i \in \Cls$ the \textbf{incidence graph} $B((F_i)_{i \in I})$ is the bipartite graph, where the two parts are given by $\bc_{i \in I} \var(F_i)$ and $I$, while there is an edge between $v$ and $i$ if $v \in \var(F_i)$. We say that \textbf{$(F_i)_{i \in I}$ is acyclic} if $B((F_i)_{i \in I})$ is acyclic (i.e., has no cycle as an (undirected) graph, or, equivalently, is a forest). A single clause-set \textbf{$F \in \Cls$ is acyclic} if $(\set{C})_{C \in F}$ is acyclic.
\end{defi}

From the family $(F_i)_{i \in I}$ of clause-sets we can derive the hypergraph $G := (\bc_{i \in I} \var(F_i), \set{\var(F_i) : i \in I})$, whose hyperedges are the variable-sets of the $F_i$. Now $(F_i)_{i \in I}$ is acyclic iff $G$ is ``Berge-acyclic'' (which just means that the bipartite incidence graph of $G$ is acyclic). The standard notion of a constraint satisfaction instance being acyclic, as defined in Subsection 2.4 in \cite{GottlobSzeider2008FPT}, is ``$\alpha$-acyclicity'' of the corresponding ``formula hypergraph'' (as with $G$, given by the variable-sets of the constraints), which is a more general notion, however since there is no automatic conversion from (sets of) clause-sets to CSP's, there is no danger of confusion here.

Since the property of the incidence graph being acyclic only depends on the occurrences of variables, if $(F_i)_{i \in I}$ is acyclic, then this is maintained by applying partial assignments and by adding new variables to each $F_i$:
\begin{lem}\label{lem:mainacyc}
  Consider an acyclic family $(F_i)_{i \in I}$ of clause-sets.
  \begin{enumerate}
  \item\label{lem:mainacyc1} For every family $(\vp_i)_{i \in I}$ of partial assignments the family $(\vp_i * F_i)_{i \in I}$ is acyclic.
  \item\label{lem:mainacyc2} Every family $(F'_i)_{i \in I}$ with $\var(F_i') \supseteq \var(F_i)$ and $(\var(F_i') \sm \var(F_i)) \cap (\var(F_j') \sm \var(F_j)) = \es$ for all $i,j \in I$, $i \ne j$, is acyclic.
  \end{enumerate}
\end{lem}

We are ready to prove that an acyclic union $\bc_{i \in I} F_i$ of clause-sets without forced assignments has itself no forced assignments. This is kind of folklore in the CSP literature (see e.g.\ \cite{Freuder1982BacktrackFreeTree,Freuder1990KTreeCSP}), using that a clause-set $F$ has no forced assignments iff $F$ considered as a single constraint is generalised arc-consistent, but to be self-contained we provide the proof. The idea is simple: any assignment to a (single) variable in some $F_i$ can be extended to a satisfying assignment $\vp_i$ of $F_i$, which sets single variables in other $F_j$, which can again be extended to satisfying assignments, and so on, and due to acyclicity never two variables are set in some $F_j$. Perhaps best to see this basic idea at work for a chain $F_1,\dots,F_m$ of clause-sets, where each $F_i$ has no forced assignments, and neighbouring clause-sets share at most one variable ($\abs{\var(F_i) \cap \var(F_{i+1})} \le 1$), while otherwise these clause-sets are variable-disjoint: then no literal $x$ is forced for $\bc_{i=1}^m F_i$, since the assignment $\pao x1$ can only affect at most two neighbouring clause-sets, for which that assignment can be extended to a satisfying assignment $\vp_0$ of them, due to them not having forced assignment (and $\var(x)$ must be the only common variable); now $\vp_0$ can be continued ``left and right'', using that assigning at most one variable in some $F_i$ can not destroy satisfiability of $F_i$.

\begin{lem}\label{lem:acyclpropc}
  Consider an acyclic family $(F_i)_{i \in I}$ of clause-sets. If no $F_i$ has forced assignments, then also $\bc_{i \in I} F_i$ has no forced assignments.
\end{lem}
\begin{prf}
We use the following simple property of acyclic graphs $G$: if $V \sse V(G)$ is a connected set of vertices and $v \in V(G) \sm V$, then there is at most one vertex in $V$ adjacent to $v$ (since otherwise there would be a cycle in $G$).

Let $F := \bc_{i \in I} F_i$ and $G := B((F_i)_{i \in I})$. Consider a literal $x \in \lit(F)$, and we show that $\pao x0 * F$ can be extended to an autarky for $F$ (the assertion then follows by Lemma \ref{lem:characnofa}); let $i_0 \in I$ with $\var(x) \in \var(F_{i_0})$. Consider a maximal $J \sse I$ with the properties:
\begin{enumerate}
\item $i_0 \in J$;
\item for $F' := \bc_{i \in J} F_i$ it is $J' := J \cup \var(F')$ connected in $G$;
\item there is a partial assignment $\vp$ with
  \begin{enumerate}
  \item $\var(\vp) = \var(F')$
  \item $\vp(x) = 0$
  \item $\vp * F' = \top$.
  \end{enumerate}
\end{enumerate}
$\set{i_0}$ fulfils these three properties (since $F_{i_0}$ has no forced assignments), and so there is such a maximal $J$. If there is no $i \in I \sm J$ adjacent to some variable in $J'$, then $\vp$ is an autarky for $F$ and we are done; so assume there is such an $i \in I \sm J$. According to the above remark, there is exactly one $v \in J'$ adjacent to $i$, that is, $\var(F_i) \cap \var(F') = \set{v}$. Since $F_j$ has no forced assignments, there is a partial assignment $\vp'$ with $\var(\vp') = \var(F_i)$, $\vp'(v) = \vp(v)$ and $\vp' * F_i = \top$. Now $\vp \cup \vp'$ satisfies $F' \cup F_i$, and thus $J \cup \set{i}$ satisfies the three conditions, contradicting the maximality of $J$. \Qed
\end{prf}

Lemma \ref{lem:acyclpropc} only depends on the boolean functions underlying the clause-sets $F_i$, and thus could be formulated more generally for boolean functions $f_i$. We obtain a sufficient criterion for the union of unit-propagation complete clause-sets to be itself unit-propagation complete:

\begin{thm}\label{thm:acylcprop}
  Consider $k \in \NNZ$ and an acyclic family $(F_i)_{i \in I}$ of clause-sets. If for all $i \in I$ we have $F_i \in \Propc_k$, then also $\bc_{i \in I} F_i \in \Propc_k$.
\end{thm}
\begin{prf}
Let $F := \bc_{i \in I} F_i$, and consider a partial assignment $\vp$ with $F' \ne \set{\bot}$ for $F' := \rk_k(\vp * F)$. We have to show that $F'$ has no forced assignments. For all $i \in I$ we have $\rk_k(\vp * F_i) \ne \set{\bot}$, and thus $\rk_k(\vp * F_i)$ has no forced assignments (since $F_i \in \Propc_k$). So $\bc_{i \in I} \rk_k(\vp * F_i)$ has no forced assignments by Lemma \ref{lem:acyclpropc}. Thus $F' = \rk_k(\bc_{i \in I} \vp * F_i) = \rk_k(\bc_{i \in I} \rk_k(\vp * F_i)) = \bc_{i \in I} \rk_k(\vp * F_i)$, whence $F'$ has no forced assignments. \Qed
\end{prf}

The conditions for $B((F_i)_{i \in I})$ being acyclic, which are relevant to us, are collected in the following lemma; they are in fact pure graph-theoretical statements on the acyclicity of bipartite graphs, but for concreteness we formulate them in terms of families of clause-sets:
\begin{lem}\label{lem:characacycl}
  Consider a family $(F_i)_{i \in I}$ of clause-sets, and let $G := B((F_i)_{i \in I})$.
  \begin{enumerate}
  \item\label{lem:characacycla} If there are $i, j \in I$, $i \ne j$, with $\abs{\var(F_i) \cap \var(F_j)} \ge 2$, then $G$ is not acyclic.
  \item\label{lem:characacyclb} Assume that for all $i, j \in I$, $i \ne j$, holds $\abs{\var(F_i) \cap \var(F_j)} \le 1$. If the ``variable-interaction graph'', which has vertex-set $I$, while there is an edge between $i, j \in I$ with $i \ne j$ if $\var(F_i) \cap \var(F_j) \ne \es$, is acyclic, then $G$ is acyclic.
  \item\label{lem:characacyclc} If there is a variable $v$, such that for $i, j \in I$, $i \ne j$, holds $\var(F_i) \cap \var(F_j) \sse \set{v}$, then $G$ is acyclic.
  \end{enumerate}
\end{lem}
\begin{prf}
For Part \ref{lem:characacycla} note that $i, j$ together with $v, w \in \var(F_i) \cap \var(F_j)$, $v \ne w$, yield a cycle (of length $4$) in $G$. For Part \ref{lem:characacyclb} assume $G$ has a cycle $C$ (which must be of even length $m \ge 4$). The case $m = 4$ is not possible, since different clause-sets have at most one common variable, and thus $m \ge 6$. Leaving out the interconnecting variables in $C$, we obtain a cycle of length $m / 2$ in the variable-interaction graph. Finally for Part \ref{lem:characacyclc} it is obvious that $G$ can not have a cycle $C$, since the length of $C$ needed to be at least $4$, which is not possible, since the only possible vertex in it would be $v$. \Qed
\end{prf}

\begin{corol}\label{cor:1acylcprop}
  Consider $k \in \NNZ$ and a family $(F_i)_{i \in I}$ of clause-sets with $F_i \in \Propc_k$ for all $i \in I$. Then each of the following conditions implies $\bc_{i \in I} F_i \in \Propc_k$:
  \begin{enumerate}
  \item\label{cor:1acylcprop1} Any two different clause-sets have at most one variable in common, and the variable-interaction graph is acyclic.
  \item\label{cor:1acylcprop2} There is a variable $v \in \Va$ with $\var(F_i) \cap \var(F_j) \sse \set{v}$ for all $i,j \in I$, $i \ne j$.
  \end{enumerate}
\end{corol}

The following examples show that the conditions of Corollary \ref{cor:1acylcprop} can not be improved in general:
\begin{examp}\label{exp:1va}
  An example for three boolean functions without forced assignments, where each pair has exactly one variable in common, while the variable-interaction graph has a cycle, and the union is unsatisfiable, is $a \oplus b = 0, \ a \oplus c = 0, \ b \oplus c = 1$. And if there are two variables in common, then also without a cycle we can obtain unsatisfiability, as $a \oplus b = 0, \ a \oplus b = 1$ shows. The latter family of two boolean functions yields also an example for a family of two clause-sets where none of them has forced assignments, while the union has (is in fact unsatisfiable). Since a hypergraph with two hyperedges is ``$\gamma$-acyclic'', in the fundamental Lemma \ref{lem:acyclpropc} we thus can not use any of the more general notions ``$\alpha$/$\beta$/$\gamma$-acyclicity'' (see \cite{Fagin1983Azyklisch} for these four basic notions of ``acyclic hypergraphs'').
\end{examp}

\section{Systems of XOR-constraints}
\label{sec:xorclausesets}

We now review the concepts of ``XOR-constraints'' and their representations via CNF-clause-sets. In Subsection \ref{sec:xorclausesetsI} we model XOR-constraints via ``XOR-clauses'' (and ``XOR-clause-sets''), and we define their semantics. And in Subsection \ref{sec:gencnfrepxor} we define ``CNF-representations'' of XOR-clause-sets, and show in Lemma \ref{lem:characimplxor} that all XOR-clauses following from an XOR-clauses are obtained by summing up the XOR-clauses.

\subsection{XOR-clause-sets}
\label{sec:xorclausesetsI}

As usual, an \textbf{XOR-constraint} (also known as ``parity constraint'') is a (boolean) constraint of the form $x_1 \oplus \dots \oplus x_n = \ve$ for literals $x_1, \dots, x_n$ and $\ve \in \set{0,1}$, where $\oplus$ is the addition in the 2-element field $\ZZ_2 = \set{0,1}$. Note that $x_1 \oplus \dots \oplus x_n = y$ is equivalent to $x_1 \oplus \dots \oplus x_n \oplus y = 0$, while $x \oplus x = 0$ and $x \oplus \ol{x} = 1$, and $0 \oplus x = x$ and $1 \oplus x = \ol{x}$. Two XOR-constraints are \emph{equivalent}, if they have exactly the same set of solutions. In this report we prefer a lightweight approach, and so we do not present a full framework for working with XOR-constraints, but we use a representation by \textbf{XOR-clauses}. These are just ordinary clauses $C \in \Cl$, but under a different interpretation, namely implicitly interpreting $C$ as the XOR-constraints $\oplus_{x \in C} = 0$. And instead of systems of XOR-constraints we just handle \textbf{XOR-clause-sets} $F$, which are sets of XOR-clauses, that is, ordinary clause-sets $F \in \Cls$ with a different interpretation. So two XOR-clauses $C, D$ are equivalent iff $\var(C) = \var(D)$ and the number of complements in $C$ has the same parity as the number of complements in $D$. That clauses are sets is justified by the commutativity of XOR, while repetition of literals is not needed due to $x \oplus x = 0$. Clashing literal pairs can be removed by $x \oplus \ol{x} = 1$ and $1 \oplus y = \ol{y}$, as long as there is still a literal left. So every XOR-constraint can be represented by an XOR-clause except of inconsistent XOR-constraints, where the simplest form is $0=1$; we can represent this by two XOR-clauses $\set{v}, \set{\ol{v}}$. In our theoretical study me might even assume that the case of an inconsistent XOR-clause-set is filtered out by preprocessing.

The appropriate theoretical background for (systems of) XOR-constraints is the theory of systems of linear equations over a field (here the two-element field). To an XOR-clause-set $F$ corresponds a system $A(F) \cdot \vec{v} = b(F)$, using ordinary matrix notation. To make this correspondence explicit we use $n := n(F)$, $m := c(F)$, $\var(F) = \set{v_1, \dots, v_n}$, and $F = \set{C_1,\dots,C_m}$. Now $F$ yields an $m \times n$ matrix $A(F)$ over $\ZZ_2$ together with a vector $b(F) \in \set{0,1}^m$, where the rows $A(F)_{i,-}$ of $A(F)$ correspond to the clauses $C_i \in F$, where a coefficient $A(F)_{i,j}$ of $v_j$ is $0$ iff $v_j \notin \var(C_i)$, and $b_i = 0$ iff the number of complementations in $C_i$ is even.
\begin{examp}\label{exp:xorcls}
  Consider $F = \set{\set{v_1,\ol{v_2}},\set{\ol{v_2},\ol{v_3}},\set{v_1,v_3}}$, where the clauses are taken in this order. Then
  \begin{displaymath}
    A(F) =
    \begin{pmatrix}
      1 & 1 & 0\\
      0 & 1 & 1\\
      1 & 0 & 1
    \end{pmatrix}, \quad b(F) =
    \begin{pmatrix}
      1\\ 0\\ 0
    \end{pmatrix}.
  \end{displaymath}
\end{examp}

\subsection{Semantical aspects}
\label{sec:gencnfrepxor}

A partial assignment $\vp \in \Pass$ satisfies an XOR-clause-set $F$ iff $\var(\vp) \supseteq \var(F)$ and for every $C \in F$ the number of $x \in C$ with $\vp(x) = 1$ is even. An XOR-clause-set $F$ implies an XOR-clause $C$ if every satisfying partial assignment $\vp$ for $F$ is also a satisfying assignment for $\set{C}$. The satisfying total assignments for an XOR-clause-set $F$ correspond 1-1 to the solutions of $A(F) \cdot \vec{v} = b$ (as elements of $\set{0,1}^n$), while implication of XOR-clauses $C$ by $F$ correspond to single equations $c \cdot \vec{v} = d$, which follow from the system, where $c$ is an $1 \times n$-matrix over $\ZZ_2$, and $d \in \ZZ_2$. Note that for every satisfiable XOR-clause-set $F$ we can compute, via computation of a row basis of $A(F)$, an equivalent XOR-clause-set $F'$ with $c(F') \le c(F)$, $n(F') \le n(F)$, and $c(F') \le n(F')$.

A \textbf{CNF-representation} of an XOR-clause-set $F \in \Cls$ is a clause-set $F' \in \Cls$ with $\var(F) \sse \var(F')$, such that the projections of the satisfying total assignments for $F'$ (as CNF-clause-set) to $\var(F)$ are precisely the satisfying (total) assignments for $F$ (as XOR-clause-set). The central question of representing XOR-clause-sets $F$ is how to obtain implied XOR-clauses $C$ from the representation $F'$; using resolution and $F'$ without auxiliary variables, it is costly in general to obtain even a single $C$ (see Section \ref{sec:transx0} for more details). How to derive a single $C$ cheaply from $F$ at the XOR-level is now discussed.

What is for CNF-clauses the resolution operation, is for XOR-clauses the addition of clauses, which corresponds to set-union, that is, from two XOR-clauses $C, D$ follows $C \cup D$. Since we do not allow clashing literals, some rule is supposed here to translate $C \cup D$ into an equivalent $E \in \Cl$ in case the two clauses are not inconsistent together. More generally, for an arbitrary XOR-clause-set $F$ we can consider the \emph{sum}, written as $\oplus F \in \Cl$, which is defined as the reduction of $\bc F$ to some clause $\oplus F := E \in \Cl$, assuming that the reduction does not end up in the situation $\set{v,\ol{v}}$ for some variable $v$ --- in this case we say that $\oplus F$ is inconsistent (which is only possible for $c(F) \ge 2$).

The following fundamental lemma translates witnessing of unsatisfiable systems of linear equations and derivation of implied equations into the language of XOR-clause-sets; it is basically a result of linear algebra, but since it might not be available in this form, we provide a proof (which is instructive anyway).
\begin{lem}\label{lem:characimplxor}
  Consider an XOR-clause-set $F \in \Cls$.
  \begin{enumerate}
  \item\label{lem:characimplxor1} $F$ is unsatisfiable if and only if there is $F' \sse F$ such that $\oplus F'$ is inconsistent.
  \item\label{lem:characimplxor2} Assume that $F$ is satisfiable. Then for all $F' \sse F$ the sum $\oplus F'$ is defined, and the set of all these clauses is modulo equivalence precisely the set of all XOR-clauses which follow from $F$.
  \end{enumerate}
\end{lem}
\begin{prf}
Obviously, if for some $F' \sse F$ we have that $\oplus F'$ is inconsistent, then $F$ is unsatisfiable, while if $F$ is satisfiable, then for every $F' \sse F$ we have that $\oplus F'$ as an XOR-clause follows from $F$. It remains to show the other directions from Part \ref{lem:characimplxor1} resp.\ \ref{lem:characimplxor2}.

We need to recall a fundamental theorem of linear algebra. Consider a field $K$ (we only need to consider $K = \ZZ_2$, but it seems that the greater generality adds lucidity here), consider $m, n \in \NNZ$, and an $m \times n$-matrix $A$ over $K$. The kernel $\ker(A) \sse K^n$ is the set of $x \in K^n$ such that $A \cdot x = 0$. We denote the rows of $A$ by $A_{1,-}, \dots A_{m,-}$, which we consider as $1 \times n$-matrices, which are identified (for convenience) with vectors in $K^n$, while the columns of $A$ are denoted by $A_{-,1}, \dots, A_{-,n}$, which are considered as $m \times 1$-matrices, and which are identified with vectors in $K^m$. The \emph{row space} of $A$ is the linear hull of the rows of $A$, denoted by $R(A) \sse K^n$, while the \emph{column space} is the linear hull of the columns of $A$, denoted by $C(A) \sse K^m$. Finally the canonical scalarproduct on $K^n$ is defined by $\inprod {x,y} := \sum_{i=1}^n x_i \cdot y_i$, and for a set $X \sse K^n$ the \emph{orthogonal complement} is $X^{\bot} := \set{y \in K^n : \inprod{x,y} = 0} \sse K^n$. Now we have the ``fundamental theorem of linear algebra'' (denoting transposition of $A$ by $\trans{A}$):\footnote{See \url{http://en.wikipedia.org/wiki/Fundamental_theorem_of_linear_algebra}, though there it is only formulated for $K = \RR$.}
\begin{eqnarray*}
  \ker(A)^{\bot} & = & R(A)\\
  \ker(\trans{A})^{\bot} & = & C(A).
\end{eqnarray*}
We show the remaining assertions first at the level of linear algebra, and then we show how to translate them to the language of XOR-clause-sets. We denote by $A' := (A,b)$ the extended matrix, which is an $m \times (n+1)$-matrix, obtained by appending $b$ as last column.

For the other direction of Part \ref{lem:characimplxor1} now assume that $A \cdot x = b$ is unsatisfiable; we show that for the vector $e_{n+1} := (0,\dots,0,1) \in K^{n+1}$ we have $e_{n+1} \in R(A')$. That $A \cdot x = b$ is unsatisfiable means that $b \notin C(A)$, that is, there is $c \in \ker(\trans{A})$ with $\inprod{c,b} \ne 0$. So $c_1 \cdot A'_{1,-} + \dots, c_m \cdot A'_{m,-}$ is a vector in $R(A')$, which is $0$ in the first $n$ components and non-zero in the last component (since this is $\inprod{c,b}$); division by the last component yields the desired result.

For the other direction of Part \ref{lem:characimplxor2} assume that $A \cdot x = b$ is satisfiable, and consider an equation $c \cdot x = d$ for some $1 \times n$-matrix $c$ and $d \in K$, which follows, that is, such that for all $x \in K^n$ holds $A x = b \Ra c x = d$. We have to show that for the vector $(c;d) \in K^{n+1}$ holds $(c;d) \in R(A')$. First we note that this holds for the case of homogeneous systems and conclusions, that is, for cases $b = 0$ and $d=0$, since then we have $c \in \ker(A)^{\bot}$, and thus $c \in R(A)$. So, introducing an additional variable $x_{n+1}$ and letting $x' = (x_1,\dots,x_n,x_{n+1})$, if we can show that the system $A' \cdot x' = 0$ implies $(c;d) \cdot x' = 0$, then we are done. So consider some $x'$ with $A' x' = 0$; we have to show that $(c;d) \cdot x' = 0$ holds.

If $x_{n+1} \ne 0$, then $x_1 A_{-,1} + \dots + x_n A_{-,n} + x_{n+1} b = 0$ is equivalent to $\frac{x_1}{-x_{n+1}} A_{-,1} + \dots +  \frac{x_n}{-x_{n+1}} A_{-,n} = b$, thus $c_1 \frac{x_1}{-x_{n+1}} + \dots + c_n \frac{x_n}{-x_{n+1}} = d$, which is equivalent to $c_1 x_1 + \dots + c_n x_n = -x_{n+1} d$, that is, $(c;d) \cdot x' = 0$.

If on the other hand $x_{n+1} = 0$ holds, then we have $A x = 0$. Since $A x = b$ is solvable, there is a solution $A x_0 = b$ (and we have $c x_0 = d$). Then we have $A (x_0+x) = b$, thus $c (x_0+x) = d$, which is equivalent to $ c x_0 + c x = d$, where $c x_0 = d$, whence $c x = 0$, that is, $(c;d) x' = 0$. This concludes the proof of the linear-algebra-formulation.

Coming finally back to the XOR-clause-sets level, for our application we have $A := A(F)$ and $b := b(F)$. We see that the row space of $A'$, considered as XOR-clauses, is precisely the set of all sums $\oplus F'$ for $F' \sse F$ (since linear combinations over $\ZZ_2$ just allow coefficients $0,1$). \Qed
\end{prf}

\section{No short AC-representations for general XOR-clause-sets}
\label{sec:noarccons}

We now show (Theorem \ref{thm:xorclsrel}), that if there were polysize AC-representations of all XOR-clause-sets, then we could translate all ``monotone span programs'' into monotone boolean circuits, which is not possible by \cite{BabaiGalWigderson1999MonoteSpanPrograms}.\footnote{See Chapter 8 of \cite{Jukna2012BooleanFunctionComplexity} for a recent introduction and overview on span programs.} We start in Subsection \ref{sec:acmon} by presenting an alternative approach to simulations by monotone circuits, as introduced in \cite{BKNW2009CircuitComplexity}. In Subsection \ref{sec:repmsp} we prove Theorem \ref{thm:xorclsrel}.

One remark on ``AC'' here: an AC-representation of a boolean function $f$ is a CNF-representation $F$ with $\phardness^{\var(f)}(F) \le 1$. We actually consider the more general condition $\hardness^{\var(f)}(F) \le 1$, and thus we obtain actually somewhat stronger results.\footnote{It seems that hardness is easier to handle in theoretical reasoning than p-hardness. Perhaps hardness is more fundamental for theoretical reasoning, with p-hardness a ``practical'' variation.} But by \cite{GwynneKullmann2013GoodRepresentationsIII} this is equivalent, when considering polynomial size (the auxiliary variables can absorb the stronger condition), and so in the general discussions here we stick to the known term ``AC''.

\subsection{AC-representations versus monotone circuits}
\label{sec:acmon}

The main result of \cite{BKNW2009CircuitComplexity} is Theorem 2, saying ``A consistency checker $f_C$ can be decomposed to a CNF of polynomial size if and only if it can be computed by a monotone circuit of polynomial size.'' We only need one side of this equivalence here, expressed more precisely by Lemma 4 there: ``Let $C_C$ be a CNF decomposition of a consistency checker $f_C$. Then, there exists a monotone circuit $S_C$ of size $O(n \abs{C_C})$ that computes $f_C$.'' Only considering the boolean case (which in any case seems the heart of the matter), and generalising / simplifying the definitions from \cite{BKNW2009CircuitComplexity}, we arrive at the following fundamental result, for which we can also provide a simpler proof. As a preparation we introduce a natural monotonisation of boolean functions:
\begin{itemize}
\item For a finite set of variable we use $\Tass(V) := \set{\vp \in \Pass : \var(\vp) = V}$ for the set of total assignments over $V$.
\item A boolean function $f(v_1,\dots,v_n)$ assigns to every total assignment $\vp$ a boolean value $f(\vp)$, and thus is a map $f: \Tass(\set{v_1,\dots,v_n}) \ra \set{0,1}$.
\item $f$ is monotone iff $(\fa\, i \in \tb 1n : v_i \le v_i') \Ra f(v_1,\dots,v_n) \le f(v_1',\dots,v_n')$.
\item We want to extend $f$ to partial assignments $\vp$, obtaining the value $0$ iff there is no total assignment $\psi \supseteq \vp$ with $f(\psi) = 1$. Furthermore, we want to obtain a monotone boolean function $\widehat{f}$, and thus setting more arguments of $\widehat{f}$ to $1$ should mean setting fewer variables of $f$ (at all).
\item For that purpose, every variable $v_i$ is replaced by two new variables $v_i', v_i''$:
  \begin{itemize}
  \item $v_i' = v_i'' = 1$ means that $v_i$ has not been assigned,
  \item $v_i' = 1$, $v_i'' = 0$ means $v_i = 0$,
  \item $v_i' = 0$, $v_i'' = 1$ means $v_i = 1$,
  \item while $v_i' = 0$, $v_i'' = 0$ means ``contradiction''.
  \end{itemize}
\item $\widehat{f}(v_1',v_1'',\dots,v_n',v_n'') = 0$ iff either
  \begin{enumerate}
  \item there is $i$ with $v_i' = v_i'' = 0$, or
  \item for the corresponding partial assignment $\vp$ (with $\var(\vp) \sse \set{v_1,\dots,v_n}$) there is no total assignment $\psi \supseteq \vp$ with $f(\psi) = 1$.
  \end{enumerate}
\item Accordingly $\widehat{f}(v_1',v_1'',\dots,v_n',v_n'') = 1$ iff there is $\psi \in \Tass(\set{v_1,\dots,v_n})$ with $f(\psi) = 1$, such that for all $i \in \tb 1n$ holds
  \begin{enumerate}
  \item if $\psi(v_i) = 0$, then $v_i' = 1$,
  \item if $\psi(v_i) = 1$, then $v_i'' = 1$.
  \end{enumerate}
\item Thus $\widehat{f}(v_1',v_1'',\dots,v_n',v_n'')$ is a monotone boolean function.
\end{itemize}

\begin{thm}\label{thm:acmono}
  Consider a boolean function $f(v_1,\dots,v_n)$ and a representation $F \in \Cls$ with $\hardness^{\set{v_1,\dots,v_n}}(F) \le 1$, that is,
  \begin{itemize}
  \item $\set{v_1,\dots,v_n} \sse \var(F)$,
  \item for every partial assignment $\vp$ with $\var(\vp) \sse \set{v_1,\dots,v_n}$, such that $f$ instantiated with $\vp$ becomes unsatisfiable, we have $\rk_1(\vp * F) = \set{\bot}$,
  \item while otherwise $\vp * F \in \Sat$.
  \end{itemize}
  From $F$ we can compute in time $O(\ell(F) \cdot n(F)^2)$ a monotone circuit $\mc{C}$ (using only binary and's and or's) which computes $\widehat{f}(v_1',v_1'',\dots,v_n',v_n'')$.
\end{thm}
\begin{prf}
In $\bot \in F$, then $\widehat{f}$ is the constant-0 function, if $F = \top$, then $\widehat{f}$ is the constant-1 function; so assume $\bot \notin F$ and $F \ne \top$. Let $N := n(F)$ and $\var(F) = \set{v_1,\dots,v_n,v_{n+1},\dots,v_N}$. Let the nodes of $\mc{C}$ be $v'_{i,j}, v''_{i,j}$ for $i \in \tb 1N$, $j \in \tb 0N$, plus one additional output-node $o$. We define $\mc{C}$ via the defining equations for its nodes.

The inputs of $\mc{C}$ are the nodes $v'_{i,0} = v'_i$, $v''_{i,0} = v''_i$ for $i \in \tb 1n$, while $v'_{i,0} = v''_{i,0} = 1$ for $i \in \tb{n+1}{N}$. The output of $\mc{C}$ is given by
\begin{displaymath}
  o = \bw_{i \in \tb 1N} v'_{i,N} \vee v''_{i,N}.
\end{displaymath}
The meaning of $v'_{i,j} = 0$ is that $v_i$ got value $1$ at stage $j$ or earlier of unit-clause propagation, while $v''_{i,j} = 0$ means that $v_i$ got value $0$ at stage $j$ or earlier of unit-clause propagation (and thus the equation for $o$ means that a contradiction was derived). For $x \in \lit(F)$ and $j \in \tb 0N$ let
\begin{displaymath}
  l(x,j) :=
  \begin{cases}
    v''_{i,j} & \text{if } x = v_i\\
    v'_{i,j} & \text{if } x = \ol{v_i}
  \end{cases}.
\end{displaymath}
For the remaining nodes with $i \in \tb 1N$ and $j \in \tb{0}{N-1}$ we have:
\begin{eqnarray*}
  v'_{i,j+1} & = & v'_{i,j} \wedge \bw_{C \in F \atop v_i \in C} \bv_{x \in C \sm \set{v_i}} l(x,j)\\
  v''_{i,j+1} & = & v''_{i,j} \wedge \bw_{C \in F \atop \ol{v_i} \in C} \bv_{x \in C \sm \set{\ol{v_i}}} l(x,j).
\end{eqnarray*}
Obviously this yields the desired meaning, and $N$ stages (layers) are enough, since at each stage of unit-clause propagation at least one new assignment is created. For each level $j$ of the defining equations we need $O(N \cdot \ell(F))$ nodes, which yields $O(N^2 \cdot \ell(F))$ nodes in the circuit altogether. \Qed
\end{prf}

\paragraph{Remarks on the monotonisation} Via $f(v_1,...,v_n) \mapsto \widehat{f}(v_1',v_2'', ..., v_n',v_n'')$ every boolean function with $n$ arguments is embedded into a monotone boolean function with $2 n$ arguments. If $f$ is given via the full truth-table, then $\widehat{f}$ can be computed in polynomial time, while if $f$ is given via an equivalent CNF $F$, then decision of $\widehat{f}(1,\dots,1)=1$ is NP-complete (since $\widehat{f}(1,\dots,1)=1$ iff $F$ is satisfiable). As pointed out by George Katsirelos\footnote{personal communication, October 2013}, there is another, related and simpler monotonisation $f'(v_1',v_2'', ..., v_n',v_n'')$, as used in \cite{Goldschlager1977MCVP}, but where now $f'$ depends on a representation of $f$ (while $\widehat{f}$ is semantically defined). Namely a deMorgan-circuit $\mc{C}$ for $f$ is taken, a monotone circuit with inputs $v_i$ and $\ol{v_i}$, and then $v_i$ is renamed to $v_i''$ and $\ol{v_i}$ to $v_i'$; in order to compare $f'$ to $\widehat{f}$, where $v_i'=v_i''=0$ means $\widehat{f}=0$, we also apply to the result additionally the conjunction of all $v_i' \vee v_i''$. We now have $\widehat{f} \le f'$, but not equality in general: Take $f = v \wedge \ol{v}$. So $f$ is constant $0$, and so is $\widehat{f}$. But $f' = (v'' \wedge v')$ (using that representation), and thus $f'$ is not constant $0$ (since $f'(1,1) = 1$).

\subsection{Characterising AC-representations}
\label{sec:charACreps}

It is instructive to see (although we do not exploit this further here in this report), that also the other direction of Theorem \ref{thm:acmono} holds, and we indeed obtain a characterisation of AC (similar to \cite{BKNW2009CircuitComplexity}):
\begin{lem}\label{lem:otherdir}
  Consider a boolean function $f$ and a monotone circuit $\mc{C}$ for $\widehat{f}$. We can construct in linear time a CNF-representation $F$ of $f$ from $\mc{C}$ with $\hardness^{\var(F)}(F) \le 1$.
\end{lem}
\begin{prf}
Let $f = f(v_1,\dots,v_n)$ and $\widehat{f} = \widehat{f}(v_1',v_1'',\dots,v_n',v_n'')$. Use the Tseitin translation of $\mc{C}$, that is, translating an or/and-node $w$ with inputs $w_1,\dots,w_m$ via using a new variable $w$ and the equivalence $w \lra (w_1 \vee \dots \vee w_m)$ resp.\ $w \lra (w_1 \wedge \dots \wedge w_m)$, to obtain a clause-set $F_0$ with $\set{v_1',v_1'',\dots,v_n',v_n''} \subset \var(F)$. Replace $v_i'$ by $\ol{v_i}$ and $v_i''$ by $v_i$ in $F_0$, remove (pseudo-)clauses containing clashing literals, add the unit-clause $\set{o}$ for the output-node $o$, and obtain $F$.

First we show that $F$ is a CNF-representation of $f$. Consider a partial assignment $\vp$ with $\var(\vp) = \set{v_1,\dots,v_n}$ and $f(\vp) = 1$. For the corresponding $\vp'$ we have $\widehat{f}(\vp') = 1$, and thus $\vp * F$ is satisfiable. While for a partial assignment $\vp$ with $\var(\vp) = \var(F)$ and $\vp * F = \top$ we have $\vp(o) = 1$, whence $f(\vp) = 1$.

It remains to show that for a partial assignment $\vp$ with $\var(\vp) \sse \set{v_1,\dots,v_n}$ and $\vp * F \in \Usat$ we have $\rk_1(\vp * F) = \set{\bot}$. Consider the corresponding $\vp'$ with $\var(\vp) = \set{v_1',v_1'',\dots,v_n',v_n''}$; we get $\widehat{f}(\vp') = 0$. To obtain that result in $\mc{C}$, variables $v_i', v_j''$ set to $1$ do not contribute. The Tseitin-equivalences contain the directions $\ol{w_1} \wedge \dots \wedge \ol{w_m} \ra \ol{w}$ (for or's) and $\ol{w_1} \vee \dots \vee \ol{w_m} \ra \ol{w}$ (for and's), and thus unit-clause propagation propagates the $0$ from the variables $v_i$ to the output-variable $o$, and then by $\set{o} \in F$ we obtain the empty clause. \Qed
\end{prf}

Instead of the full Tseitin translation, which uses equivalences, one can also use the reduced translation in the proof of Lemma \ref{lem:otherdir}, using the implications $w \ra (w_1 \vee \dots \vee w_m)$ and $w \ra (w_1 \wedge \dots \wedge w_m)$. That leads to clauses of the type $\set{\ol{w},w_1,\dots,w_m}$, $\set{\ol{w},w_i}$, and thus $F_0$ here is a pure dual Horn clause-set (has exactly one negative literal in each clause). Corollary 3 in \cite{BKNW2009CircuitComplexity} makes a similar statement: ``Let $C_C$ be a CNF decomposition of a consistency checker $f_C$. The variables of $C_C$ can be renamed to that each clause has exactly one negative literal.'' In our more general context (for the boolean case), this renamability to pure dual Horn clause-sets does not hold for arbitrary representations $F$ of boolean functions $f$ of relative hardness $1$, but only after first constructing a monotone circuit $\mc{C}$ from $F$ by Theorem \ref{thm:acmono}, and then transforming $\mc{C}$ into a representation of relative hardness $1$ by Lemma \ref{lem:otherdir} (and ignoring negative literals for variables in $f$, and also ignoring the added unit-clause).

\begin{examp}\label{exp:monohorn}
  We consider the boolean function $f = (a \vee b \vee c) \wedge (\neg a \vee \neg b \vee \neg c)$ and its representation $F_0 := \primec_0(f) = \set{\set{a,b,c},\set{\ol{a},\ol{b},\ol{c}}}$. This representation can not be renamed into a (dual) Horn clause-set, but of course after removal of negative literals it is trivially dual Horn. We note that $F_0$ itself can be used here to obtain a monotone circuit $\mc{C}$ computing $\widehat{f}$, since $F_0$ is of hardness $0$. Furthermore we can use gates of arbitrary fan-in, since we allow clauses of arbitrary size. So the monotone circuit (in fact, a monotone formula here) for $\widehat{f}$ is
  \begin{displaymath}
    \mc{C} = ((a'' \vee b'' \vee c'') \wedge (a' \vee b' \vee c')) \wedge (a' \vee a'') \wedge (b' \vee b'') \wedge (c' \vee c').
  \end{displaymath}
  Using the reduced Tseitin translation, we obtain $w_1 \ra (a'' \vee b'' \vee c'')$, $w_2 \ra (a' \vee b' \vee c')$, $w_3 \ra (a' \vee a'')$, $w_4 \ra (b' \vee b'')$, $w_5 \ra (c' \vee c'')$, and $o \ra w_1 \wedge w_2 \wedge w_3 \wedge w_4 \wedge w_5$ from $\mc{C}$, and thus we get the pure dual Horn clause-set
  \begin{multline*}
    F_0' = \set{\set{\ol{w_1},a'',b'',c''},\set{\ol{w_2},a',b',c'},\\
      \set{\ol{w_3},a',a''},\set{\ol{w_4},b',b''},\set{\ol{w_5},c',c''},\\
      \set{\ol{o},w_1},\dots,\set{\ol{o},w_5}}.
  \end{multline*}
  Finally we get $F = \set{\set{\ol{w_1},a,b,c}, \set{\ol{w_2},\ol{a},\ol{b},\ol{c}}, \set{\ol{o},w_1},\dots,\set{\ol{o},w_5}, \set{o}}$.
\end{examp}

We conclude by characterising AC-representations:
\begin{corol}\label{cor:otherdir}
  A sequence $(f_n)_{n \in \NN}$ of boolean functions has a CNF-representation $(F_n)_{n \in \NN}$ with $\hardness^{\var(f_n)}(F_n) \le 1$ and $\ell(F_n) = n^{O(1)}$ if and only if the sequence $(\widehat{f_n})_{n \in \NN}$ can be computed by monotone circuits of size polynomial in $n$.
\end{corol}

By \cite{GwynneKullmann2013GoodRepresentationsIII} the condition ``$\hardness^{\var(f_n)}(F_n) \le 1$'' in Corollary \ref{cor:otherdir} can be replaced by ``$h^{\var(f_n)}(F_n) \le k$'' for any $h \in \set{\hardness,\phardness,\whardness}$ and any fixed $k \in \NN$.

\subsection{Representing monotone span programs}
\label{sec:repmsp}

We are ready to prove that there are no short AC-representations of XOR-systems (recall Subsection \ref{sec:introlowb} for an overview on the proof idea).
\begin{thm}\label{thm:xorclsrel}
  There is no polynomial $p$ such that for all XOR-clause-sets $F \in \Cls$ there is a CNF-representation $F' \in \Cls$ with $\ell(F') \le p(\ell(F))$ and $\hardness^{\var(F)}(F') \le 1$.
\end{thm}
\begin{prf}
We consider representations of monotone boolean functions
\begin{displaymath}
  f: \set{0,1}^n \ra \set{0,1}
\end{displaymath}
by ``monotone span programs'' (msp's):
\begin{itemize}
\item The input variables are given by $x_1,\dots,x_n$.
\item Additionally $m \in \NNZ$ boolean variables $y_1,\dots,y_m$ can be used, where $m$ is the dimension, which we can also be taken as the size of the span program.
\item For each $i \in \tb 1n$ there is a linear system $A_i \cdot y = b_i$ over $\ZZ_2$, where $A_i$ is an $m_i \times m$ matrix with $m_i \le m$, and $b_i \in \set{0,1}^{m_i}$.
\item For a total assignment $\vp$, i.e., $\vp \in \Pass$ with $\var(\vp) = \set{x_1,\dots,x_n}$, the value $f(\vp)$ is $0$ if and only if the linear systems given by $\vp(x_i) = 0$ together are unsatisfiable, that is,
  \begin{displaymath}
    f(\vp) = 0 \iff \setb{y \in \set{0,1}^m \mb \fa\, i \in \tb 1n : \vp(x_i)=0 \Ra A_i \cdot y = b_i} = \es.
  \end{displaymath}
\end{itemize}
W.l.o.g.\ we assume that each system $A_i \cdot y = b_i$ is satisfiable.

Consider for each $i \in \tb 1n$ an XOR-clause-set $A_i' \in \Cls$ representing $A_i \cdot y = b_i$; so $\var(A_i') \supseteq \set{y_1,\dots,y_m}$, where, as always, new variables for different $A_i'$ are used, that is, for $i \ne j$ we have $(\var(A_i') \cap \var(A_j')) \sm \set{y_1,\dots,y_m} = \es$. We use the process $\doping: \Cls \ra \Cls$ of ``doping'', as introduced in \cite{GwynneKullmann2013GoodRepresentationsIII}, where $\doping(F)$ is obtained from $F$ by adding to each clause a new variable. Let $A_i'' := \doping(A_i')$, where the doping variables for different $i$ do not clash; we denote them (altogether) by $z_1,\dots,z_N$. Let $F := \bc_{i=1}^n A_i''$. Consider a CNF-representation $F'$ of the XOR-clause-set $F$. We have
\begin{displaymath}
  f(\vp)=0 \iff \vp' * F' \in \Usat,
\end{displaymath}
where $\vp'$ is a partial assignment with $\vp'$ assigning only doping variables $z_j$, namely if $\vp(x_i) = 0$, then all the doping variables used in $\doping(A_i')$ are set to $0$, while if $\vp(x_i) = 1$, then nothing is assigned here. The reason is that by setting the doping variables to $0$ we obtain the original system $A_i \cdot y = b_i$, while by leaving them in, this system becomes satisfiable whatever the assignments to the $y$-variables are.

Now assume that we have $\hardness^{\set{z_1,\dots,z_N}}(F') \le 1$. By Theorem \ref{thm:acmono} we obtain from $F'$ a monotone circuit $\mc{C}$ (using only and's and or's) of size polynomial in $\ell(F')$ with input variables $z_1',z_1'',\dots,z_N',z_N''$, where
\begin{itemize}
\item $z_j' = z_j'' = 1$ means that $z_j$ has not been assigned,
\item $z_j' = 1$, $z_j'' = 0$ means $z_j = 0$,
\item $z_j' = 0$, $z_j'' = 1$ means $z_j = 1$,
\item while $z_j' = 0$, $z_j'' = 0$ means ``contradiction'' (where the output of $\mc{C}$ is $0$).
\end{itemize}
The value of $\mc{C}$ is $0$ iff the corresponding partial assignment applied to $F'$ yields an unsatisfiable clause-set. In $\mc{C}$ we now replace the inputs $z_j',z_j''$ by inputs $x_i$, which in case of $x_i = 0$ sets $z_j'=1$, $z_j'' = 0$ for all related $j$, while in case of $x_i = 1$ all related $z_j',z_j''$ are set to $1$.\footnote{In other words, all $z_j'$ are set to $1$, while $z_j'' = x_i$ for the $j$ related to $i$.} This is now a monotone circuit computing $f$. By \cite{BabaiGalWigderson1999MonoteSpanPrograms}, Theorem 1.1, thus it is not possible that $F'$ is of polynomial size in $F$. \Qed
\end{prf}

In \cite{GwynneKullmann2013GoodRepresentationsIII} we show that in the (unrestricted) presence of auxiliary variables even w-hardness boils down, modulo polytime computations, to $\Propc_1 = \Propc$ under the relative condition:
\begin{corol}\label{cor:xorcls}
  XOR-clause-sets do not have good representations with bounded w-hardness, not even when using relative w-hardness. That is, there is no $k \in \NNZ$ and no polynomial $p(x)$ such that for all XOR-clause-sets $F \in \Cls$ there exists a CNF-representation $F' \in \Cls$ with $\ell(F') \le p(\ell(F))$ and $\whardness^{\var(F)}(F') \le k$.
\end{corol}

\section{The most basic translation $X_0$}
\label{sec:transx0}

After having shown that there is no ``small'' AC-representation of arbitrary XOR-clause-sets $F$, the task is to find ``good'' CNF-representations for special $F$. First we consider $c(F)=1$, that is, a single XOR-clause $C$, to which we often refer as ``$x_1 \oplus \dots \oplus x_n = 0$''. There is precisely one equivalent clause-set, i.e., there is exactly one representation without auxiliary variables, namely
\begin{displaymath}
  \bmm{X_0(C)} := \primec_0(x_1 \oplus \dots \oplus x_n = 0) \in \Urefc_0,
\end{displaymath}
the set of prime implicates of the underlying boolean function, which is unique since the prime implicates are not resolvable. $X_0(C)$ has $2^{n-1}$ clauses for $n \ge 1$ (while for $n=0$ we have $X_0(C) = \top$), the full clauses (containing all variables) over $\set{\var(x_1),\dots,\var(x_n)}$, where the parity of the number of complementations is different from the parity of the number of complementations in $C$. Note that for two XOR-clauses $C, D$ we have $X_0(C) = X_0(D)$ iff $C, D$ are equivalent. By definition we have $X_0(C) \in \Urefc_0$.

More generally, we define $X_0: \Cls \ra \Cls$, where the input is interpreted as XOR-clause-set and the output as CNF-clause-set, by $\bmm{X_0(F)} := \bc_{C \in F} X_0(C)$. By Theorem \ref{thm:acylcprop} and Lemma \ref{lem:mainacyc}, Part \ref{lem:mainacyc2}:
\begin{lem}\label{lem:suffx0pc}
  If $F \in \Cls$ is acyclic, then $X_0(F) \in \Propc$.
\end{lem}

In the rest of this section we consider $X_0(F)$ for unsatisfiable XOR-clause-sets $F$. These cases can be handled by preprocessing, but nevertheless they are instructive, and they have been at the heart of lower bounds for the resolution calculus from the beginnings. Ignoring the size of the obtained representation, the following simple example shows that $X_0(\set{C,D})$ for $C, D \in \Cl$ in general has high w-hardness.
\begin{examp}\label{exp:2xor0}
  For $n \in \NN$ and (different) variables $v_1,\dots,v_n$ consider the system
  \begin{eqnarray*}
    v_1 \oplus v_2 \oplus \dots \oplus v_n &=& 0 \\
    v_1 \oplus v_2 \oplus \dots \oplus \ol{v_n} &=& 0,
  \end{eqnarray*}
  that is, consider the XOR-clauses $C_1 := \set{v_1,\dots,v_n}$ and $C_2 := \set{v_1,\dots,v_{n-1},\ol{v_n}}$. Then $X_0(\set{C_1,C_2})$ is the clause-set with all $2^n$ full clauses over $\set{v_1,\dots,v_n}$, and thus $\hardness(X_0(\set{C_1,C_2})) = \whardness(X_0(\set{C_1,C_2})) = n$ (Lemma 3.18 in \cite{Ku99b}).
\end{examp}

An early and very influential example of (hard) unsatisfiable clause-sets are the ``Tseitin formulas'' introduced in \cite{Ts68}, which are defined as follows. Consider a general graph $G = (V,E,\eta)$, that is, $V$ is the set of vertices, $E$ is the set of edge-labels, while $\eta: E \ra \binom V2$ maps every edge-label to some $e \sse V(G)$ with $1 \le \abs{e} \le 2$. The special conditions are that $E$ is a clause, and there is a ``charge'' $\rho: V \ra \set{0,1}$. In order that we only have to deal with XOR-clauses, we forbid the case that an isolated vertex can have charge $1$ (that would lead to ``$0 = 1$''). For every vertex $w \in V(G)$ the XOR-clause $C_w$ now expresses
\begin{displaymath}
  \oplus_{x \in E(G), w \in \eta(x)} x = \rho(w),
\end{displaymath}
that is, the XOR over all literal-edges incident with $w$ is $\rho(w)$. Let $T_0(G,v,\rho) := \set{C_w : w \in V(G)}$ be the XOR-clause-set derived from $G$. Then $T_0(G,v,\rho)$ is unsatisfiable if $G$ has no loops and  $\oplus_{w \in V(G)} \rho(w) = 1$, since $\oplus_{w \in V(G)} \oplus_{x \in E(G), w \in \eta(x)} x = 0$, due to every edge occurring precisely twice in the sum. Finally the \textbf{Tseitin clause-set} is $T(G,v,\rho) := X_0(T_0(G,v,\rho))$.

\begin{examp}\label{exp:2xor0T}
  For an XOR-clause $C$ we have $X_0(C) = T(B_C)$, where $B_C$ is the bouquet with one vertex $C$, which has charge $0$, and the literals of $C$ as edges.

  The two XOR-clauses $C_1, C_2$ from Example \ref{exp:2xor0} are the two XOR-clauses associated with the dipole $D_n$, which is the general graph with two vertices and the variables $v_1,\dots,v_n$ as edges connecting these two vertices, where the first vertex gets charge $0$ and the second gets charge $1$. And thus $X_0(\set{C_1,C_2}) = T(D_n)$.
\end{examp}

In \cite{Ts68} an exponential lower bound for regular resolution refutations of (special) Tseitin clause-sets was shown, and thus unsatisfiable Tseitin clause-sets in general have high hardness. This was extended in \cite{Urq87} to full resolution, and thus unsatisfiable Tseitin clause-sets in general also have high w-hardness. In the following we refine $X_0: \Cls \ra \Cls$ in various ways, by first transforming an XOR-clause-set $F$ into another XOR-clause-set $F'$ representing $F$, and then using $X_0(F')$.

\section{The standard translation $X_1$}
\label{sec:transx1}

If the XOR-clause-set $F$ contains long clauses, then $X_0(F)$ is not feasible, and the XOR-clauses of $F$ have to be broken up into short clauses, which we consider now. As we have defined how a CNF-clause-set can represent an XOR-clause-set, we can define that an XOR-clause-set $F'$ represents an XOR-clause-set $F$, namely if the satisfying assignments of $F'$ projected to the variables of $F$ are precisely the satisfying assignments of $F$.

\begin{defi}\label{def:1softxor}
  Consider an XOR-clause $C = \set{x_1,\dots,x_n} \in \Cl$. The \textbf{natural splitting} of $C$ is the XOR-clause-set $F'$ obtained as follows, using $n := \abs{C}$:
  \begin{itemize}
  \item If $n \le 2$, then $F' := \set{C}$.
  \item Otherwise choose pairwise different new variables $y_2, \dots, y_{n-1} \in \Va \sm \var(C)$, and let
    \begin{displaymath}
      F' := \set{x_1 \oplus x_2 = y_2} \cup \set{y_{i-1} \oplus x_i = y_i}_{i \in \tb 3{n-1}} \cup \set{y_{n-1} \oplus x_n = 0}
    \end{displaymath}
    (i.e., $F' = \set{\set{x_1,x_2,y_2}} \cup \set{\set{y_{i-1},x_i,y_{i}}}_{i \in \tb 3{n-1}} \cup \set{\set{y_{n-1},x_n}}$).
  \end{itemize}
  Then $F'$ as XOR-clause-set is a representation of $\set{C}$. Let $\bmm{X_1(C)} := X_0(F')$.
\end{defi}
For $C \in \Cl$ and $n := \abs{C}$ we have for $F := X_1(C)$:
\begin{itemize}
\item If $n \le 2$, then $n(F) = c(F) = n$, and $\ell(F) = 2^{n-1} \cdot n$.
\item Otherwise $n(F)=2n-2$, $c(F)=4n-6$ and $\ell(F) = 12n-20$.
\end{itemize}

\begin{examp}\label{exp:xortrans}
  For $n = 3$ we get
  \begin{displaymath}
    X_1(C) = \setb{ \underbrace{\set{x_1,x_2,\ol{y_2}}, \set{x_1,\ol{x_2},y_2}, \set{\ol{x_1},x_2, y_2}, \set{\ol{x_1},\ol{x_2},\ol{y_2}}}_{\bmm{x_1 \oplus x_2 = y_2}},\underbrace{\set{y_2,\ol{x_3}}, \set{\ol{y_2},x_3} }_{\bmm{y_2 \oplus x_3 = 0}} }.
  \end{displaymath}
\end{examp}

In Example \ref{exp:2xor0T} we have seen how to obtain $X_0(C)$ as a Tseitin clause-sets (via a bouquet). Also $X_1(C)$ can be obtained as a Tseitin clause-set, as the following example shows.

\begin{examp}\label{exp:x1ts}
  Consider a clause $C = \set{x_1,\dots,x_n}$. As we have seen, $X_1(C)$ for $n \le 2$ is the Tseitin clause-set for the bouquet given by the literals $x_1,\dots,x_n$ (in both cases having one vertex $v_1$). Now the general graph $G_n$ for $n \ge 3$, such that $X_1(C)$ is the Tseitin clause-set for $G_n$, is obtained recursively from $G_{n-1}$ by adding one new vertex $v_n$, which has two incident edges, namely $y_{n-1}$ connected to the last vertex $v_{n-1}$ added (where $v_2 := v_1$) and the loop $x_n$; all charges are $0$.
\end{examp}

Corollary \ref{cor:1acylcprop}, Part \ref{cor:1acylcprop2}, applies to $F'$ from Definition \ref{def:1softxor}, and thus:
\begin{lem}\label{lem:1softxor}
  For $C \in \Cl$ we have $X_1(C) \in \Propc$.
\end{lem}
We define $X_1: \Cls \ra \Pcls{3}$, where the input is interpreted as XOR-clause-set and the output as CNF-clause-set, by $X_1(F) := \bc_{C \in F} X_1(C)$ for $F \in \Cls$, where some choice for the new variables is used, so that the new variables for different XOR-clauses do not overlap. By Theorem \ref{thm:acylcprop}, Lemma \ref{lem:1softxor} and Lemma \ref{lem:mainacyc}, Part \ref{lem:mainacyc2} we get:
\begin{thm}\label{thm:suffx1pc}
  If $F \in \Cls$ is acyclic, then $X_1(F) \in \Propc$.
\end{thm}
A precursor to Theorem \ref{thm:suffx1pc} is found in Theorem 1 of \cite{LaitinenJunttilaNiemelae2012Parity}, where it is stated that tree-like XOR clause-sets are ``UP-deducible'', which is precisely the assertion that for acyclic $F \in \Cls$ the representation $X_1(F)$ is AC. As mentioned in \cite{LaitinenJunttilaNiemelae2012Parity}, such XOR clause-sets have good applications, with 61 out of 474 SAT benchmarks from SAT competition 2005 to 2011 containing only tree-like systems of XOR equations.

The question is now how much Theorem \ref{thm:suffx1pc} can be extended. In Section \ref{sec:transtxor} we will see that $X_0(F)$ and $X_1(F)$ have high hardness in general, even for $c(F) = 2$. But we will also see that appropriate preprocessing of the XOR-clause-set improves the yield of $X_1$. Of course, from Corollary \ref{cor:xorcls} we know that even under the relative condition there is no general solution.

\section{AC for XOR-clause-sets is fpt in the number of equations}
\label{sec:transarbxor}

As we have shown, for a CNF-clause-set $F$ the computation of an equivalent $F'$ with $\hardness(F') = 0$ is fixed-parameter tractable (fpt) in the number $c(F)$ of clauses, and so we obtain that computation of the representation $F'$ with (absolute) hardness $0$ of an XOR-clause-set $F$ with fixed maximal clause-length $k$ is fpt in $c(F)$:
\begin{lem}\label{lem:fptkc}
  Consider a constant $k \in \NNZ$ and an XOR-clause-set $F \in \Pcls{k}$. The (unique) CNF-representation $F'$ of $F$ without auxiliary variables, consisting of precisely the CNF-prime-implicates of the XOR-clause-set $F$ (and thus $\hardness(F') = 0$), has size $c(F') \le 16^{k \cdot c(F)}$, and can be computed in time $O(\ell(F) \cdot 4096^{k \cdot c(F)})$.
\end{lem}
\begin{prf}
We have $c(X_1(F)) \le 4 k \cdot c(F)$, and thus by Lemma \ref{lem:primcec} we can compute a CNF-representation $F_0 \in \Urefc_0$ of $X_1(F)$ of size $c(F_0) \le 2^{4 k \cdot c(F)}$ in time $O(\ell(F) \cdot 2^{12 k \cdot c(F)}$. Selecting the clauses $C \in F_0$ with $\var(C) \sse \var(F)$ we obtain $F'$. \Qed
\end{prf}

It seems an interesting feature of the proof of Lemma \ref{lem:fptkc}, that we cannot use directly $X_0(F)$, since that would yield more clauses ($c(X_0(F)) \le 2^{k-1} c(F)$), but the detour via $X_1(F)$ (containing auxiliary variables) seems needed. We obtain that computing a CNF-representation with (absolute) hardness $0$ of an XOR-clause-set $F \in \Pcls{k}$ is fpt in $c(F)$ for each fixed $k$. When allowing CNF-representations with relative p-hardness $1$ (an AC-representation), then we obtain fpt in the parameter $c(F)$ for arbitrary $F \in \Cls$ (now also the constants involved are small):
\begin{thm}\label{thm:relxorcnfp}
  Consider a satisfiable XOR-clause-set $F \in \Cls$. Let $F^* := \set{\oplus F' : F' \sse F} \in \Cls$ (recall Lemma \ref{lem:characimplxor}); $F^*$ is computable in time $O(\ell(F) \cdot 2^{c(F)})$ (while $c(F^*) \le 2^{c(F)}$). Then $\bmm{X^*(F)} := X_1(F^*)$ is a CNF-representation of $F$ with $\phardness^{\var(F)}(X_1(F^*)) \le 1$.
\end{thm}
\begin{prf}
Consider some partial assignment $\vp$ with $\var(\vp) \sse \var(F)$, let $F' := \rk_1(\vp * F^*)$, and assume there is a forced literal $x \in \lit(F')$ for $F'$. Then the XOR-clause $C := \set{y \in \Lit : \vp(y) = 0} \cup \set{\ol{x}}$ follows from $F$. By Lemma \ref{lem:characimplxor} there is $F' \sse F$ with $\oplus F' = C$ modulo equivalence of XOR-clauses. So we have (modulo equivalence) $X_1(C) \sse F^*$, where due to $X_1(C) \in \Propc$ (Lemma \ref{lem:1softxor}) the forced literal $x$ for $\vp * X_1(C)$ is set by $\rk_1$, contradicting the assumption. \Qed
\end{prf}

Theorem 4 in \cite{LaitinenJunttilaNiemelae2013Parity} yields the weaker bound  $O(4^{n(F)})$ for the number of clauses in an AC-representation of $F$ (note that w.l.o.g.\ $c(F) \le n(F)$). The following example shows the representation of Theorem \ref{thm:relxorcnfp} is not in $\Propc$.
\begin{examp}\label{exp:arccfptnpc}
  Consider the XOR-clauses $C := \set{a,b,c,d}$ and $D := \set{a,b,c,e}$, where $a,\dots,e$ are different variables, and let $F := \set{C,D}$. Then we have $F^* = \set{C,D,\set{d,e},\bot}$ (where $\bot$ can be removed, also in general), and $\hardness(X_1(F^*)) = 2$.

Let $y^C_2,y^C_3$ and $y^D_2,y^D_3$ be the new variables in $C$ resp.\ $D$ (so, semantically, $y^C_2 = a \oplus b = y^D_2$ and $y^C_3 = y^C_2 \oplus c$,  $y^D_3 = y^D_2 \oplus c$). To see that $\hardness(X_1(F^*)) \ge 2$, observe that $F' := \pab{y^C_2 \ra 0, y^D_2 \ra 1} * X_1(F^*)$ is unsatisfiable (that is, forcing $a \oplus b = 0 \und a \oplus b = 1$), but all clauses in $F'$ are of size $2$ (no unit-clauses).

Considering the upper-bound, the only way to make $X_1(F^*)$ unsatisfiable without immediately yielding the empty-clause, is (essentially) to set one of the new variables $y_2, y_3$ in each $X_1(C), X_1(D)$ to contradictory values. So consider the different possibilities. If $y^C_2$ is set to $b \in \set{0,1}$ and $y^D_2$ is set to $1 - b$, then $F' := \pab{y^C_2 \ra b, y^D_2 \ra 1 - b} * X_1(F^*) \in \Pcls{2}$, and hence $\hardness(F') \le 2$. Otherwise, if $y^C_3$ is set to $b \in \set{0,1}$ and $y^D_3$ is set to $1-b$, then unit-clause propagation forces $d$ and $e$ in $F' := \pab{y^C_3 \ra b, y^D_3 \ra 1 - b} * X_1(F^*)$ to opposing values, and hence creates the empty-clause in $X_1(\set{d,e})$.
\end{examp}

In Lemma \ref{lem:unbhd2xc} we will see that in fact absolute hardness $\hardness(F^*)$ even just for $c(F)=2$ is unbounded. On the other hand, in Conjecture \ref{con:relxorcnfp} we state our belief that we can strengthen Theorem \ref{thm:relxorcnfp} by also establishing absolute (p-)hardness $1$. We now turn to the problem of understanding and refining the basic translation $X_1$ for two clauses.

\section{Translating two XOR-clauses}
\label{sec:transtxor}

For an XOR-clause-set $F$ with $c(F) \le 1$ we have $X_1(F) \in \Propc$, which is a perfect representation. We are now considering in detail the case of $c(F) = 2$. By Theorem \ref{thm:relxorcnfp} we can consider $F^* = \set{C,D,\oplus\set{C,D}}$, and obtain the CNF-representation $X_1(F^*)$ of relative p-hardness $1$. But as Example \ref{exp:arccfptnpc} shows, absolute p-hardness is larger than $1$, and Lemma \ref{lem:unbhd2xc} indeed shows that (absolute) p-hardness is unbounded.

\subsection{In $\Propc$}

With more intelligence, we can provide a representation in $\Propc$ as follows; note that an XOR-clause-set $\set{C,D}$ is unsatisfiable iff $\abs{C \cap \ol{D}}$ is odd and $\var(C) = \var(D)$.
\begin{thm}\label{thm:2xorshared}
  Consider two XOR-clauses $C, D \in \Cl$. We assume the following (to simplify the presentation), using $V := \var(C) \cap \var(D)$:
  \begin{itemize}
  \item $\abs{V} \ge 2$.
  \item $\abs{C} > \abs{V}$ and $\abs{D} > \abs{V}$.
  \item Thus w.l.o.g.\ $\abs{C \cap D} = \abs{V}$.
  \end{itemize}
  Let $I := C \cap D$,
  \begin{enumerate}
  \item Choose $s \in \Va \sm \var(\set{C,D})$, and let $I' := I \cup \set{s}$.
  \item Let $C' := (C \sm I) \cup \set{s}$ and $D' := (D \sm I) \cup \set{s}$.
  \item Now $\set{I',C',D'}$ is an XOR-clause-set which represents the XOR-clause-set $\set{C,D}$.
  \end{enumerate}
  Let $\bmm{X_2(C,D)} := X_1(\set{I', C', D'})$.
  Then $X_2(C,D) \in \Propc$ is a CNF-representation of the XOR-clause-set $\set{C,D}$.
\end{thm}
\begin{prf}
That $\set{I',C',D'}$ represents $\set{C,D}$ is obvious, since $s$ is the sum of the common part. Corollary \ref{cor:1acylcprop}, Part \ref{cor:1acylcprop2}, applies to $\set{I',C',D'}$ (the only common variable is $s$), and thus we get $X_2(C,D) \in \Propc$. \Qed
\end{prf}

We believe that this method can be generalised to more than two clauses:
\begin{conj}\label{con:relxorcnfp}
  We can combine a generalisation of Theorem \ref{thm:2xorshared} with Theorem \ref{thm:relxorcnfp} and obtain $X_*: \Cls \ra \Propc$, which computes for an XOR-clause-set $F \in \Cls$ a CNF-representation $X_*(F)$ in time $2^{O(c(F))} \cdot \ell(F)^{O(1)}$.
\end{conj}
The stronger Conjecture \ref{con:relxorcnfptw} replaces $c(F)$ by the treewidth $\twidth(F) \in \NNZ$ of the incidence graph of $F$; since for the complete bipartite graphs $K_{m,n}$, $m,n \in \NNZ$, we have $\twidth(K_{m,n}) = \min(m,n)$, and removing edges does not increase the treewidth, we have $\twidth(F) \le c(F)$.

\subsection{In $\Wrefc_3$}
\label{sec:2xwc3}

We now turn to the analysis of the ``naked'' translation $X_1(\set{C,D})$ for two XOR-clauses $C, D$. We can show in \cite{BeyersdorffGwynneKullmann2013PHPER}, that these representations have high relative hardness:
\begin{lem}\label{lem:hdtwoxor} \cite{BeyersdorffGwynneKullmann2013PHPER}
  For two XOR-clauses $C, D$ except of trivial cases holds
  \begin{displaymath}
    \hardness(X_1(\set{C,D})) =
    \hardness^{\var(\set{C,D})}(X_1(\set{C,D})) = \max(1,\abs{\var(C)
      \cap \var(D)}).
  \end{displaymath}
\end{lem}
It follows that the addition of derived clauses is not sufficient to keep hardness low, when considering the application of partial assignments to the auxiliary variables:
\begin{lem}\label{lem:unbhd2xc} \cite{BeyersdorffGwynneKullmann2013PHPER}
  For two XOR-clauses $C, D$, $\hardness(X_1(\set{C,D}^*))$ is arbitrarily large (using $F^*$ as defined in Theorem \ref{thm:relxorcnfp}).
\end{lem}

So the distance from AC for $X_1(\set{C,D})$ is as large as possible, and this is provably the worst translation from the three considered. However, still it has merits, namely (absolute) w-hardness is in fact low, which we show now. First we consider the worst-case, the unsatisfiable case where $C, D$ coincide except of one flipped literal; this case can be produced from the general case by application of partial assignments. Though we won't use it here, it is instructive to obtain these clause-sets via the Tseitin method (compare Examples \ref{exp:2xor0T}, \ref{exp:x1ts}):
\begin{examp}\label{exp:twoxorts}
  Consider the XOR-clauses $C_1, C_2$ from Example \ref{exp:2xor0}. Let $T_n := X_1(\set{C_1,C_2})$. We can obtain $T_n$ also as a Tseitin clause-set, where the principle of construction of the underlying general graph should become clear from the following example for $n=4$:
  
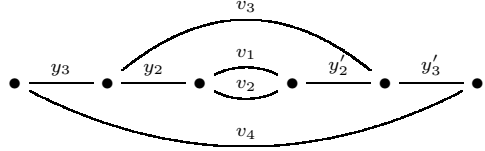
\begin{displaymath}
    \xymatrix {
      {\bullet} \aru[r]^{y_3} \aru@/_2pc/[rrrrr]^{v_4} & {\bullet} \aru[r]^{y_2} \aru@/^2pc/[rrr]^{v_3} & {\bullet} \aru@/^/[r]^{v_1} \aru@/_/[r]^{v_2} & {\bullet} \aru[r]^{y_2'} & {\bullet} \aru[r]^{y_3'} & {\bullet} &
    }
  \end{displaymath}
  All vertices have charge $0$ except of the rightmost vertex.
\end{examp}
In \cite{BeyersdorffGwynneKullmann2013PHPER} we show $\hardness(T_n) = n$, and thus these clause-sets are very hard regarding tree-resolution. We show now that for dag-resolution the refutation is very easy.
\begin{thm}\label{thm:2xor}
  Consider $T_n$ from Example \ref{exp:twoxorts}. We have:
  \begin{enumerate}
  \item $n(T_n) = 2 \cdot (2n-2) - n = 3n - 4$ for $n \ge 2$.
  \item $c(T_n) = 8 n - 12$ for $n \ge 2$.
  \item $\ell(T_n) = 24 n - 40$ for $n \ge 2$.
  \item $T_n \in \Usat \cap \Pcls{3}$.
  \item For $n \ge 3$ holds $\whardness(T_n) = \wid(T_n) = 3$.
  \item There exists a resolution refutation using altogether $18 n - 29$ clauses.
  \end{enumerate}
\end{thm}
\begin{prf}
To show the lower bound for $\whardness$, consider the closure $\widehat{T_n}$ of $T_n$ under $2$-resolution. The binary clauses in $T_n$ are exactly $\primec_0(y_{n-1} = x_n \und \ol{y'_{n-1}} = x_n)$. The resolution of these binary clauses with ternary clauses in $T_n$ allows the corresponding substitutions ($y_{n-1} = x_n = \ol{y'_{n-1}}$) to be made in (other) clauses containing those variables, but this does not introduce any further clauses of size $\le 2$. Therefore, $\widehat{T_n}$ contains only clauses of size $\ge 2$, so $\whardness(T_n) \ge 3$. To show $\wid(T_n) \le 3$, we construct a resolution refutation.

From
\begin{eqnarray*}
  \primec_0(y_{n-1} \oplus x_n = 0) &=& \set{\set{\ol{y_{n-1}},x_n}, \set{y_{n-1},\ol{x_n}}} \\
  \primec_0(y'_{n-1} \oplus \ol{x_n} = 0) &=& \set{\set{\ol{y'_{n-1}},\ol{x_n}},\set{y'_{n-1},x_n}},
\end{eqnarray*}
via $2$-resolution we derive $\primec_0(y_{n-1} = \ol{y'_{n-1}}) = \set{\set{\ol{y_{n-1}},\ol{y'_{n-1}}}, \set{y_{n-1},y'_{n-1}}}$:
\begin{prooftree}
  \AxiomC{$\set{\ol{y_{n-1}},x_n}$}
  \AxiomC{$\set{\ol{y'_{n-1}},\ol{x_n}}$}
  \BinaryInfC{$\set{\ol{y_{n-1}},\ol{y'_{n-1}}}$}
  \AxiomC{$\set{y_{n-1},\ol{x_n}}$}
  \AxiomC{$\set{y'_{n-1},x_n}$}
  \BinaryInfC{$\set{y_{n-1},y'_{n-1}}$}
  \alwaysNoLine\BinaryInfC{~}
\end{prooftree}

From
\begin{eqnarray*}
  \primec_0(y_{i-1} \oplus x_i = y_i) &=& \set{\underbrace{\set{\ol{y_{i-1}},\ol{x_i},\ol{y_i}}}_{\bmm{C_1}}, \underbrace{\set{\ol{y_{i-1}},x_i,y_i}}_{\bmm{C_2}}, \underbrace{\set{y_{i-1},\ol{x_i}, y_i}}_{\bmm{C_3}}, \underbrace{\set{y_{i-1},x_i,\ol{y_i}}}_{\bmm{C_4}}} \\
  \primec_0(y'_{i-1} \oplus x_i = y'_i) &=& \{ \underbrace{\set{\ol{y'_{i-1}}, \ol{x_i}, \ol{y'_i}}}_{\bmm{D_1}}, \underbrace{\set{\ol{y'_{i-1}}, x_i, y'_i}}_{\bmm{D_2}}, \underbrace{\set{y'_{i-1}, \ol{x_i}, y'_i}}_{\bmm{D_3}}, \underbrace{\set{y'_{i-1}, x_i, \ol{y'_i}}}_{\bmm{D_4}} \}\\
  \primec_0(y_i = \ol{y'_i}) &=& \set{\underbrace{\set{\ol{y_i},\ol{y'_i}}}_{\bmm{E_1}},\underbrace{\set{y_i,y'_i}}_{\bmm{E_2}}}
\end{eqnarray*}
we derive $\primec_0(y_{i-1} = \ol{y'_{i-1}}) = \set{\set{\ol{y'_{i-1}}, \ol{y_{i-1}}}, \set{y'_{i-1}, y_{i-1}}}$:
\begin{prooftree}\small
  \def\defaultHypSeparation{\hskip .1in}
  \AxiomC{$C_1$}
  \AxiomC{$E_2$}
  \BinaryInfC{$\set{\ol{y_{i-1}},\ol{x_i},y'_i}$}
  \AxiomC{$D_2$}
  \BinaryInfC{$\set{\ol{y'_{i-1}},\ol{y_{i-1}},y'_i}$}
  \AxiomC{$C_2$}
  \AxiomC{$E_1$}
  \BinaryInfC{$\set{\ol{y_{i-1}},x_i,\ol{y'_i}}$}
  \AxiomC{$D_1$}
  \BinaryInfC{$\set{\ol{y'_{i-1}},\ol{y_{i-1}},\ol{y'_i}}$}
  \BinaryInfC{\bmm{\set{\ol{y'_{i-1}}, \ol{y_{i-1}}}}}
  \AxiomC{$C_3$}
  \AxiomC{$E_1$}
  \BinaryInfC{$\set{y_{i-1},\ol{x_i},\ol{y'_i}}$}
  \AxiomC{$D_4$}
  \BinaryInfC{$\set{y'_{i-1},y_{i-1},\ol{y'_i}}$}
  \AxiomC{$C_4$}
  \AxiomC{$E_2$}
  \BinaryInfC{$\set{y_{i-1},x_i,y'_i}$}
  \AxiomC{$D_3$}
  \BinaryInfC{$\set{y'_{i-1},y_{i-1},y'_i}$}
  \BinaryInfC{\bmm{\set{y'_{i-1}, y_{i-1}}}}
  \alwaysNoLine\BinaryInfC{~}
\end{prooftree}

Hence, by induction on $n$, we derive $\primec_0(y_2 = \ol{y'_2})$. We conclude: From
\begin{eqnarray*}
  \primec_0(x_1 \oplus x_2 = y_2) &=& \set{\underbrace{\set{\ol{x_1},\ol{x_2},\ol{y_2}}}_{\bmm{C_1}}, \underbrace{\set{\ol{x_1},x_2,y_2}}_{\bmm{C_2}}, \underbrace{\set{x_1,\ol{x_2}, y_2}}_{\bmm{C_3}}, \underbrace{\set{x_1,x_2,\ol{y_2}}}_{\bmm{C_4}}} \\
  \primec_0(x_1 \oplus x_2 = y'_2) &=& \set{\underbrace{\set{\ol{x_1},\ol{x_2},\ol{y'_2}}}_{\bmm{D_1}}, \underbrace{\set{\ol{x_1},x_2,y'_2}}_{\bmm{D_2}},\underbrace{\set{x_1,\ol{x_2}, y'_2}}_{\bmm{D_3}}, \underbrace{\set{x_1,x_2,\ol{y'_2}}}_{\bmm{D_4}}} \\
  \primec_0(y_2 = \ol{y'_2}) &=& \set{\underbrace{\set{\ol{y_2},\ol{y'_2}}}_{\bmm{E_1}},\underbrace{\set{y_2,y'_2}}_{\bmm{E_2}}}
\end{eqnarray*}
we derive $\bot$:
\begin{prooftree}
  \def\defaultHypSeparation{\hskip .1in}
  \AxiomC{$C_1$}
  \AxiomC{$E_2$}
  \BinaryInfC{$\set{\ol{x_1},\ol{x_2},y'_2}$}
  \AxiomC{$D_1$}
  \BinaryInfC{$\set{\ol{x_1},\ol{x_2}}$}
  \AxiomC{$C_2$}
  \AxiomC{$E_1$}
  \BinaryInfC{$\set{\ol{x_1},x_2,\ol{y'_2}}$}
  \AxiomC{$D_2$}
  \BinaryInfC{$\set{\ol{x_1},x_2}$}
  \BinaryInfC{$\set{\ol{x_1}}$}
  \AxiomC{$D_3$}
  \AxiomC{$C_3$}
  \AxiomC{$E_1$}
  \BinaryInfC{$\set{x_1,\ol{x_2},\ol{y'_2}}$}
  \BinaryInfC{$\set{x_1,\ol{x_2}}$}
  \AxiomC{$D_4$}
  \AxiomC{$C_4$}
  \AxiomC{$E_2$}
  \BinaryInfC{$\set{x_1,x_2,y'_2}$}
  \BinaryInfC{$\set{x_1,x_2}$}
  \BinaryInfC{$\set{x_1}$}
  \BinaryInfC{$\bot$}
\end{prooftree}

The number of clauses in this refutation (which uses only clauses of length at most $3$) altogether is
\begin{enumerate}
\item $8 n - 12$ clauses from $T_n$.
\item $2$ clauses from the derivation of $\primec_0(y_{n-1} = \ol{y'_{n-1}})$.
\item $(n-3) \cdot 10$ clauses from $(n-3)$ induction steps.
\item $11$ clauses in the final refutation in step.
\end{enumerate}
So in total, the resolution proof is of size $18 n - 29$. \Qed
\end{prf}

For arbitrary XOR-clauses $C, D$ the worst-case for instantiation of $X_1(\set{C,D})$ happens when we get the situation of $T_n$ above, and thus:
\begin{corol}\label{cor:whd2xor}
  For an XOR-clause-set $F$ with $c(F) \le 2$ holds $\wid(X_1(F)) \le 3$.
\end{corol}

\subsection{Discussion}
\label{sec:2dis}

To summarise, there are three levels of representing two XOR-clauses $C, D$:
\begin{description}
\item[Low w-hardness] For $F_1 := X_1(\set{C,D})$ we have low width (thus low w-hardness), namely $\wid(F_1) \le 3$, but high relative hardness.
\item[Relative p-hardness 1] For $F_2 := X_1(C,D,\oplus\set{C,D})$ we have relative p-hardness $1$, but high absolute hardness.
\item[P-hardness 1] For $F_3 := X_2(C,D)$ we have absolute p-hardness $1$.
\end{description}

The most drastic cure seems to resolve Conjecture \ref{con:relxorcnfp}, so that for a constant number of clauses we can reach p-hardness $1$ in polynomial time. While the most lazy approach is to do nothing, relying on the w-hardness not growing too much:
\begin{conj}\label{con:whardc}
  There is a function $\alpha: \NNZ \ra \NNZ$ such that for all XOR-clause-sets $F \in \Cls$ holds $\whardness(X_1(F)) \le \alpha(c(F))$.
\end{conj}
We know $\alpha(0) = 0$ and $\alpha(1) = 3$. Since $X_1(F)$ has clause-length at most $3$, we have $\wid(X_1(F)) \le \whardness(X_1(F)) + \max(\whardness(X_1(F)),3)$, and so we could have required as well $\wid(X_1(F)) \le \alpha(c(F))$ in Conjecture \ref{con:whardc}.

\section{Conclusion and open problems}
\label{sec:open}

We investigated ``good'' SAT representations $F'$ of systems of linear equations over $\set{0,1}$, handled via XOR-clause-sets $F$. We showed that even under the most generous measurement of quality of $F'$, relative w-hardness, i.e., $\whardness^{\var(F)}(F') \le k$, there do not exist $F'$ of polynomial size in general for constant $k$. Then we considered the possibilities of computing $F'$ of relative p-hardness $1$ ($\phardness^{\var(F)}(F') \le 1$), i.e., AC, or absolute p-hardness $1$ ($\phardness(F') \le 1$), that is, $F' \in \Propc$. The methodology in general is to transform $F$ into another XOR-clause-set $G$, and then to use $F' = X_0(G)$ (translating every XOR-clause into the unique equivalent CNF-clause-set). By adding to $F$ all sums of subsets of $F$ (as XOR-clauses) we obtain AC, where the computation is fixed-parameter tractable in the number of clauses of $F$. Our remaining endeavours are about obtaining $F' \in \Propc$. We achieved this for two cases for $F$:
\begin{itemize}
\item $F$ is acyclic;
\item $c(F) \le 2$.
\end{itemize}
In the first case $X_1(F)$ does the job, where $X_1$ just splits clauses up, so that $X_0$ only has to handle XOR-clauses of length at most $3$. While the second case is handled by $X_2(F)$, which factors out the common part of the two XOR-clauses.

The case $c(F) = 2$ we considered more closely, and showed that even with just using $X_1(F)$ we get low w-hardness, however relative hardness is high. Using the general method to obtain AC, we then obtain relative p-hardness $1$, however absolute hardness is still high. Finally, via the translation $X_2$ we get absolute p-hardness $1$.

\subsection{Open problems and future research directions}

Theorem \ref{thm:acylcprop} is a basic general tool for obtaining clause-sets in $\Propc$, based on acyclic graphs. This should be generalised by considering treewidth and related notions. Applications in Lemma \ref{lem:suffx0pc}, Theorem \ref{thm:suffx1pc}, and Theorem \ref{thm:2xorshared} yield methods to obtain representations of XOR-clause-sets in $\Propc$, while Conjecture \ref{con:relxorcnfp} says, that computing a representation in $\Propc$ should be fixed-parameter tractable in the number of XOR-clauses. More generally, we conjecture to have fixed-parameter tractability in the treewidth of the incidence graph (strengthening Conjecture \ref{con:relxorcnfp}):
\begin{conj}\label{con:relxorcnfptw}
  There exists $X^*: \Cls \ra \Propc$, which computes for an XOR-clause-set $F \in \Cls$ a CNF-representation $X^*(F)$ in time $2^{O(\twidth(F))} \cdot \ell(F)^{O(1)}$, where $\twidth(F)$ is the treewidth of the incidence graph of $F$.
\end{conj}
Theorem 7 in \cite{LaitinenJunttilaNiemelae2013ParityE} shows a weaker form of Conjecture \ref{con:relxorcnfptw}, where instead of the incidence graph the variable-interaction graph (or ``primal graph''; recall Lemma \ref{lem:characacycl}) is considered, and where instead of (absolute) propagation-completeness only AC is achieved. Altogether there seem to be interesting general theorems and methods waiting to be discovered, targeting representations in $\Propc$.

Considering the negative results, the main question for Theorem \ref{thm:xorclsrel} and Corollary \ref{cor:xorcls} is to obtain sharp bounds on the size of shortest representations $F'$ with $\phardness^{\var(F)}(F') \le k$ resp.\ $\hardness^{\var(F)}(F') \le k$ resp.\ $\whardness^{\var(F)}(F') \le k$ for fixed $k$.

Finally, in Subsection \ref{sec:2xwc3} we started a complexity analysis (hardness analysis) of XOR-representations (extended by Conjecture \ref{con:whardc}). While in Examples \ref{exp:2xor0T}, \ref{exp:x1ts} and \ref{exp:twoxorts} we gave connections to Tseitin clause-sets. Combining these streams would likely offer valuable insights.

In this article we have only considered representations of XOR-clause-sets via CNF-clause-sets. Extending the CNF-mechanism however is also a necessary avenue (as shown by Theorem \ref{thm:xorclsrel} and Corollary \ref{cor:xorcls}), and we present a theoretical perspective in the following subsection (based on a semantic perspective, not on a proof-theoretic perspective as the DPLL(XOR)-framework introduced in \cite{Laitinen2010DPLLParity}).

\subsection{Hard boolean functions handled by oracles}
\label{sec:conclhard}

By Corollary \ref{cor:xorcls} we know that systems of XOR-clauses (affine equations) have no good representation, even when just considering AC. To overcome these limitations, the theory started here has to be generalised via the use of oracles as developed in \cite{Ku99b,Ku00g}, and further discussed in Subsection 9.4 of \cite{GwynneKullmann2012Slur,GwynneKullmann2012SlurJ}. The point of these oracles, which are just sets $\mc{U} \sse \Usat$ of unsatisfiable clause-sets stable under application of partial assignments, is to discover hard \emph{unsatisfiable} (sub-)instances (typically in polynomial time). We obtain relativised hierarchies $\Urefc_k(U)$, $\Propc_k(U)$, $\Wrefc_k(U)$, which are defined as before, with the only change that $\Urefc_0(U) \cap \Usat = \Propc_0(U) \cap \Usat = \Wrefc_0(U) \cap \Usat := U$.

The use of such oracles is conceptually simpler than the current integration of SAT solvers and methods from linear algebra (see Subsection \ref{sec:introspecreas}). Recall the simple CNF-representation $X_1: \Cls \ra \Cls$ of XOR-clause-sets. In Theorem \ref{thm:2xor} we have seen that already for two clauses this is a bad translation (at least from the hardness-perspective). Now let $U_{X_1}$ be the set of unsatisfiable $\vp * X_1(F)$ for $F \in \Cls$ and $\vp \in \Pass$ (it is not hard to see that $U_{X_1}$ is decidable in polynomial time). Then we have $X_1: \Cls \ra \Urefc_0(U_{X_1})$.

An important aspect of the theory to be developed must be the usefulness of the representation (with oracles) in context, that is, as a ``constraint'' in a bigger problem: a boolean function $f$ represented by a clause-set $F$ is typically contained in $F^* \supset F$, where $F^*$ is the SAT problem to be solved (containing also other constraints). One approach is to require from the oracle also stability under addition of clauses, as we have it already for the resolution-based reductions like $\rk_k$, so that the (relativised) reductions $\rk_k^{\mc{U}}$ can always run on the whole clause-set (an instantiation of $F^*$). However for example for the oracle mentioned below, based on semidefinite programming, this would be prohibitively expensive. And for some oracles, like detection of minimally unsatisfiable clause-sets of a given deficiency, the problems would turn from polytime to NP-hard in this way (\cite{FKS00,BueningZhao2002MUsubsets}). Furthermore, that we have some (representation of a) constraint which would benefit for example from some XOR-oracle, does not mean that in other parts of the SAT-problems that oracle will also be of help. So in many cases it is better to restrict the application of the oracle $\mc{U}$ to that subset $F \subset F^*$, where the oracle is actually required to achieve the desired hardness.

Another example of a current barrier is given by the satisfiable pigeonhole clause-sets $\php^m_m$, which have variables $p_{i,j}$ for $i, j \in \tb 1m$, and where the satisfying assignments correspond precisely to the permutations of $\tb 1m$. The question is about ``good'' representations. In \cite{BeyersdorffGwynneKullmann2013PHPER} we show $\hardness(\php^m_m) = \whardness(\php^m_m) = m-1$, and so the (standard representation) $\php^m_m \in \Cls$ itself is not a good representation (it is small, but has high w-hardness). Moreover, as shown in \cite{BeyersdorffGwynneKullmann2013PHPER}, from Theorem \ref{thm:acmono} it follows that $\php^m_m$ has no polysize AC-representation (or, more generally, of bounded relative w-hardness) at all. So again, here oracles are needed; see Subsection 9.4 of \cite{GwynneKullmann2012Slur,GwynneKullmann2012SlurJ} for a proposal of an interesting oracle based on semidefinite programming (with potentially good stability properties).

\bibliographystyle{plain}

\newcommand{\noopsort}[1]{}

\end{document}